\documentclass[aip,
amsmath,amssymb,
 reprint,
]{revtex4-1}

\usepackage{graphicx} 
\usepackage{dcolumn}
\usepackage{bm}

\usepackage[utf8]{inputenc}
\usepackage[T1]{fontenc}
\usepackage{mathptmx}
\usepackage{color}
\usepackage{amsmath}
\usepackage{hyphenat}
\usepackage{hyperref}
\usepackage{color}
\usepackage{amstext}
\usepackage{braket}

\usepackage{natbib}

\usepackage{hyperref}
\hypersetup{
    colorlinks=true,
    linkcolor=blue,
    filecolor=magenta,      
    urlcolor=blue,
}

\begin{document}
\preprint{AIP/123-QED}

\title{Substitutional Si impurities in monolayer hexagonal boron nitride}

\author{Mohammad Reza Ahmadpour Monazam}
\email{mohammad.monazam@univie.ac.at}
\affiliation{University of Vienna, Faculty of Physics, Boltzmanngasse 5, A-1090, Vienna, Austria}
\author{Ursula Ludacka}
\affiliation{University of Vienna, Faculty of Physics, Boltzmanngasse 5, A-1090, Vienna, Austria}
\author{Hannu-Pekka Komsa}
\affiliation{Department of Applied Physics, Aalto University, P.O. Box 11100, 00076 Aalto, Finland}
\author{Jani Kotakoski}
\affiliation{University of Vienna, Faculty of Physics, Boltzmanngasse 5, A-1090, Vienna, Austria}

\date{\today}

\begin{abstract}
We report the observation of substitutional silicon atoms in single-layer hexagonal boron nitride (h-BN) using aberration corrected scanning transmission electron microscopy (STEM). The images reveal silicon atoms exclusively filling boron vacancies. Density functional theory (DFT) is used to study the energetics, structure and properties of the experimentally observed structure. The formation energies reveal  Si$_\mathrm{B}^{+1}$ as the most stable configuration. In this case, silicon atom elevates by 0.66~\text{\AA} out of the lattice with unoccupied defect levels in the electronic band gap above the Fermi level. Our results unequivocally show that heteroatoms can be incorporated into the h-BN lattice opening way for applications ranging from single-atom catalysis to atomically precise magnetic structures.
\end{abstract}

\maketitle

The study of two-dimensional (2D) materials has since the introduction of graphene~\cite{Novoselov2004} opened an active research field in material science. Graphene was quickly followed by other 2D material such as hexagonal boron nitride (h-BN)~\cite{Pacile2008} and transition metal dichalcogenides~\cite{Wang2012}, that in contrast to graphene exhibit an electronic band gap. Among 2D materials, h-BN has attracted attention due to its high thermal and chemical stability, high thermal conductivity and low dielectric constant, besides its wide electronic band gap~\cite{Cassabois2016}, although due to its honeycomb-like arrangement of \textit{$sp^2$}-hybridized structure that is almost a perfect match to graphene, it is most often regarded just a suitable substrate for graphene-based applications~\cite{Geim2013}.

Defects play a crucial role in semiconductors in determining the applicability of the material. For example, vacancies and impurities change the electronic and optoelectronic properties by adding localized defect levels into their band gaps~\cite{Oba2018, Kapetanakis2016, Kotakoski2011} providing charge carrier traps and/or combination centers~\cite{Hu2018}. Although silicon is the most often encountered impurity atom in graphene samples~\cite{Zhou2012,Ramasse2013,Jalili2018}, it has not been until now observed in h-BN. In contrast, both oxygen and carbon have been found in single-layer h-BN, probably due to electron beam damage, during a STEM experiment~\cite{Krivanek2010}. On the one hand, Si impurities in h-BN could be interesting from material engineering point of view, especially for applications in electronics \cite{Meyer2009,Kim2018,Murata2013,Majety2013}, quantum computing \cite{Sajid2018} and spintronic devices \cite{Tang2010,Asshoff2018,Wang2018}. On the other hand, they could be detrimental when h-BN is used as a gate dielectric in field-effect transistors. Previous theoretical studies have shown that silicon substitution in boron vacancy is more stable than substitution in nitrogen vacancy~\cite{Liu2014, Mapasha22016, Mapasha2016}. Both $+1$ and $-1$ have also been considered in addition to the neutral state~\cite{Mapasha2016}. However, the number of valence electrons and the position of the defect levels have been ignored.

In this Letter, we show the atomic resolution STEM images of substitutional silicon impurities in h-BN with DFT calculations and image simulations revealing the details of the atomic configuration. The formation energies for each of the possible silicon substitutions (in boron, nitrogen and double vacancy) are calculated with different charge states. In accordance with the experiments that show exclusively impurity atoms in the B lattice site, our simulations reveal Si$_\mathrm{B}^{+1}$ as the configuration with the lowest formation energy. Our results demonstrate the possibility of incorporating heteroatoms into h-BN opening way for atomic-scale engineering of the material for applications.

Single-layer h-BN suffers from significant electron beam damage in transmission electron microscopy experiments~\cite{Meyer2009,Jin2009,Kotakoski2010}, caused by a combination of knock-on processes and ionization damage. The vacancies created during electron irradiation tend to grow fast into triangular holes, but can also be filled by atoms from the ubiquitous hydrocarbon-based contamination covering most samples, as presumably happened in the study presented in Ref.~\cite{Krivanek2010}. In contrast to such non-intrinsic defects created during the experiment, we discuss here intrinsic impurity atoms that were found in the samples after preparation with no additional processing and only a minimal electron dose.

\begin{figure}[ht!]
    \centering
	\includegraphics[width=.40\textwidth]{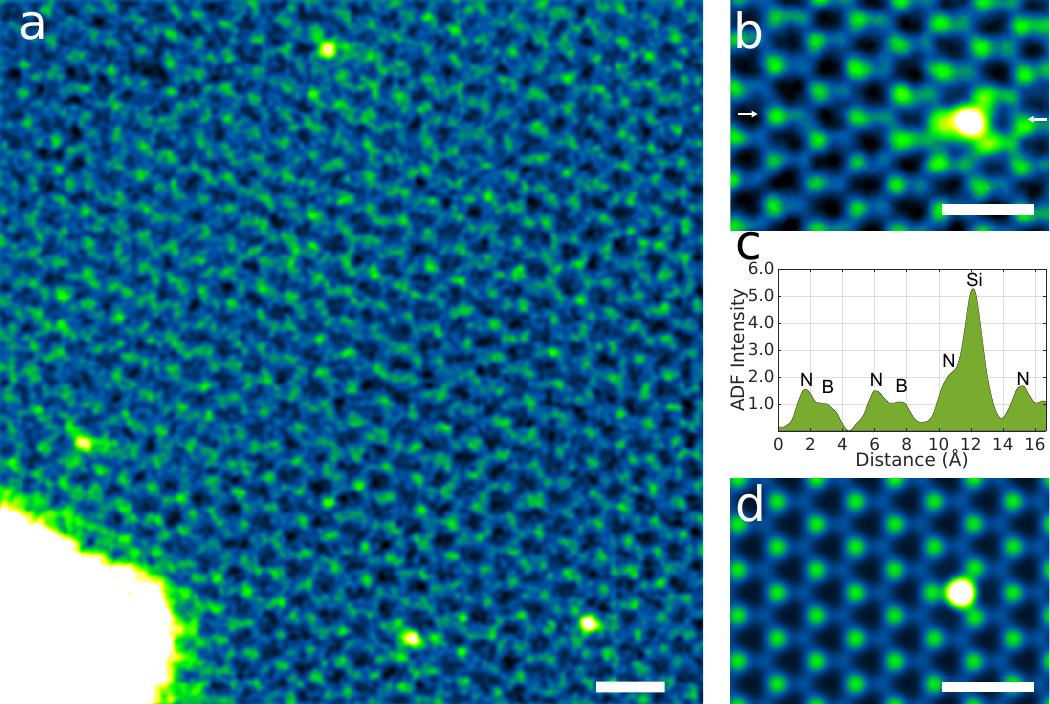}
\caption{\label{MAADF} {\bf Silicon substitution at boron vacancy in h-BN.} (a) Experimental MAADF image of multiple Si atoms in h-BN (bright points) recorded with a relatively low dose to avoid electron-beam damage. (b) A close-up MAADF image of Si$_\mathrm{B}$ (the brightest atom). N and B atoms can be distinguished based on their contrast (brighter and darker, respectively). (c) Line profile showing the intensities between the arrows in panel b. (d) Simulated MAADF image of Si$_\mathrm{B}^{+1}$ (QSTEM package~\cite{Koch2002}). The scale bars are 0.5~nm.} 
\end{figure}

We point out that unlike most h-BN samples, that are prepared via mechanical exfoliation, ours were grown via chemical vapor deposition (Graphene Laboratories, Inc.), which may explain why Si impurities in h-BN have not been reported until now. The samples were directly transferred onto golden transmission electron microscopy grids with perforated amorphous carbon membrane (Quantifoil\textregistered) without a polymer. The copper was etched in FeCl over night and the samples were baked in vacuum at 150$^\circ$C overnight before being inserted into the microscope. Fig.~\ref{MAADF}a shows an atomically resolved medium angle annular dark field (MAADF) image of the suspended h-BN membrane with four silicon impurity atoms in the h-BN lattice. All images were acquired with the Nion UltraSTEM 100 microscope~\cite{krivanek_electron_2008} in Vienna at 60~keV with near-ultrahigh vacuum conditions at the objective area (pressure $<10^{-9}$~mbar)~\cite{leuthner_scanning_2018}. The beam convergence semiangle was 30~mrad and the MAADF detector angular range was 60-200~mrad. Typical beam current of the device is on the order of 30~pA.

The silicon atom can be easily distinguished by its higher contrast as compared to boron and nitrogen~\cite{Krivanek2010}. In all images, Si impurities are on boron sites (see Fig.~\ref{MAADF}b). The image intensities are shown in Fig.~\ref{MAADF}c along the path marked by arrows in Fig.~\ref{MAADF}b, normalized to the intensity corresponding to a single boron atom. Intensity ratio between Si and B is expected to be $(14/5)^{1.64} = 5.41$~\cite{Krivanek2010}, close to our experimental value of 5.25. Although the substituted silicon atom is stable enough to allow its repeated scanning, it can not sustain the electron dose required for electron energy loss spectroscopy~\cite{ramasse_probing_2013} or energy-dispersive X-ray spectroscopy to independently confirm the chemical identity of the individual impurity atoms. Since the hydrocarbon-based contamination covering much of the sample also contains silicon, spectra recorded over larger areas would remain inconclusive. Fig.~\ref{MAADF}d shows a simulated MAADF image of the relaxed structure (+1 charge state) with a good agreement between the projected Si-N distances between the nearest nitrogen atoms and the impurity in calibrated experimental and simulated images. For both cases the projected Si-N distance is around 1.55~$\text{\AA}$. The distance between neighboring nitrogen atoms shows an increase from 2.51~$\text{\AA}$ in pristine h-BN to 2.74~$\text{\AA}$ around the impurity, consistent with the optimized model structure.

The typical hydrocarbon-based contamination that covers practically all graphene and h-BN samples contains a large number of silicon atoms. In fact, it is possible to dope such structures with Si simply by creating vacancies into them at elevated temperatures \cite{Inani2019}. Hence, it is natural to assume that this is also the source of those Si atoms found in our h-BN samples.

We turn to density functional theory (DFT) calculations (as implemented in the Vienna ab initio simulation package (VASP)~\cite{Kresse1993}) to try to understand why they are exclusively found in the B lattice sites.  The electron exchange and correlation was treated by Perdew-Burke-Ernzerdorf (PBE) functional~\cite{Perdew1996}. The total energy of the system was calculated via the pseudopotential-momentum-space formalism using projector-augmented-wave (PAW) method~\cite{Kresse1999}. The Kohn-Sham wavefunctions are expanded over plane-wave basis sets with the kinetic energy cut off set to 525~eV. Converged locally optimized configurations and formation energies were found for a supercell of $8\times8\times1$. The interlayer vacuum space of 43.46~\text{\AA} was selected according to "special vacuum" proposed in Refs.~\cite{komsa2014, Komsa2018}. The Brillouin-Zone integration was done over a $\Gamma$-centered $5\times5\times1$ k-point mesh. The damped molecular dynamics method was used to optimize the ionic degrees of freedom until residual forces were below 0.01~eV/$\text{\AA}$. We point out that although it is known that the band gaps calculated using PBE underestimate the true band gap of semiconductors, it yields formation energies similar to those calculated with the HSE (Heyd-Scuseria-Ernzerhof)~\cite{Heyd2003} formalism~\cite{Berseneva2011}. This allows rescaling of the electron chemical potential of PBE calculations using the difference in the band gap obtained from the two methods leading to a significant saving in the computational cost. Here, HSE (HSE06 functional with 0.25 fraction of exchange~\cite{Aliaksandr2006}) was used to calculate the band gap of bulk h-BN (5.72 eV as compared to 4.48 eV calculated with PBE) for this purpose.

\begin{figure}[hb!]
	\includegraphics[width=.40\textwidth]{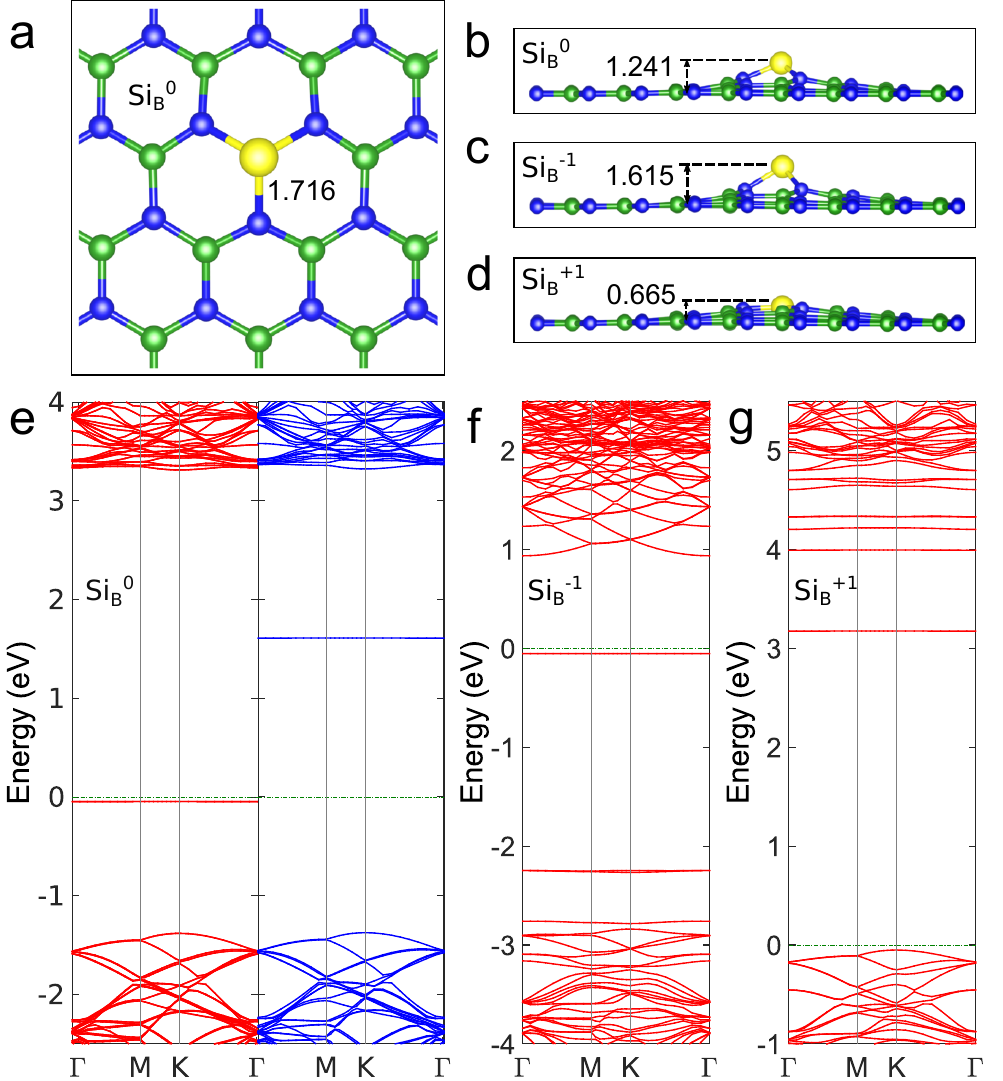}
	\caption{\label{SiB-BS} {\bf Atomic and band structure of Si in boron vacancy.} (a) Top view of locally optimized Si$_\mathrm{B}$ with Si-N bond length shown. Side view of (b) Si$_\mathrm{B}^0$, (c) Si$_\mathrm{B}^{-1}$ and (d) Si$_\mathrm{B}^{+1}$ including the estimated elevation of silicon with respect to the h-BN plane. The PBE-calculated spin-polarized band structure of (e) Si$_\mathrm{B}^0$ with two defect levels in the band gap, and spin-unpolarized band structures of (f) Si$_\mathrm{B}^{-1}$ and (g) Si$_\mathrm{B}^{+1}$.}
\end{figure}

For each possible impurity site in h-BN, we first structurally optimize the structure, and then calculate its electronic properties and formation energy. The calculations are repeated for different charge states. Fig.~\ref{SiB-BS}a-b show the top and side view of the structures for neutral state of Si$_\mathrm{B}$. The N-Si bond length is found to be $\sim{1.72}$ \text{\AA}, which is significantly longer than $\sim{1.45}$ $\text{\AA}$ between B and N in the pristine structure and the 1.55~\text{\AA} measured from the experimental images. The silicon atom rises 1.241~$\text{\AA}$ above the h-BN plane. The side view of the structures corresponding to different charge states are shown in Fig.~\ref{SiB-BS}c-d. As for the +1 charge state, the N-Si bond length drops to 1.63~$\text{\AA}$. For -1 charge state the bond length becomes 1.82~\text{\AA}. Simply taking into account the projected distance between silicon and neighboring nitrogen atoms is not conclusive enough to estimate the charge state of the defect. The simulated MAADF images show the projected Si-N distance within $1.53-1.55$~\text{\AA}, which are very close to the calibrated distances of experimental data. Negative charge elevates Si further away from the negatively charged N neighbors, whereas positive charge has the opposite effect.

The band structures of silicon substitution in boron vacancies at different charge states are plotted in Fig.~\ref{SiB-BS}e-g. The Fermi level is set to zero. The red (blue) lines correspond to spin up (down) band structure. For the neutral state, Si substitution adds two defect states within the band gap, where only one of the defect levels (spin down) is empty. Therefore, Si$_\mathrm{B}$ has only two expected charge states (-1 and +1); adding further electrons or holes to the structure leads to electrons in the nearly-free-electron (NFE) state of conduction band minimum (CBM) or holes at the valence band maximum (VBM)~\cite{Berseneva2011,Huang2011}. This NFE state is estimated to be 2~$\text{\AA}$ away from the h-BN plane~\cite{Huang2012}. In the +1 and -1 charge states the system becomes spin-unpolarized. An interesting case is the band structure of Si$_\mathrm{B}^{-1}$ (Fig.~\ref{SiB-BS}f). Here, the defect level is so close to the Fermi level that added electron is almost free and should be easy to extract.

We calculate the defect formation energy using the supercell method~\cite{Walle1989}. Here, the formation energy at a charge state $q$ is defined as
\begin{multline*}
E^f[X^q] = E_{tot}[X^q] - E_{tot}[host]  \\
        - \sum n^i\mu_i + q[E_F + E_{VBM}] + E_{corr},
\end{multline*}
where $E_{tot}[X^q]$ is the total energy of the supercell containing a defect or impurity $X$, $E_{tot}[host]$ is the total energy for equivalent supercell of perfect crystal. $n_i$ are the number of atoms which are added ($n_i > 0$) or removed ($n_i < 0$) from the supercell and $\mu_i$ are the chemical potentials of the constituent atoms $i$. The formation energy is expressed as a function of electron chemical potential (i.e., Fermi energy $E_F$ with respect to the valence band maximum of the pristine structure).  $E_{corr}$ corresponds to all spurious electrostatic corrections due to employing the supercell method. We calculate the chemical potential for boron and nitrogen atoms as the total energy of $\beta$-rhombohedral boron (per atom), containing 106 atoms per unit cell \cite{Setten2007}, and half of the chemical potentials of the nitrogen molecule (N$_2$). The chemical potential of silicon has been calculated from total energy of bulk silicon (per atom).

\begin{figure}[htbp]
	\includegraphics[width=0.40\textwidth]{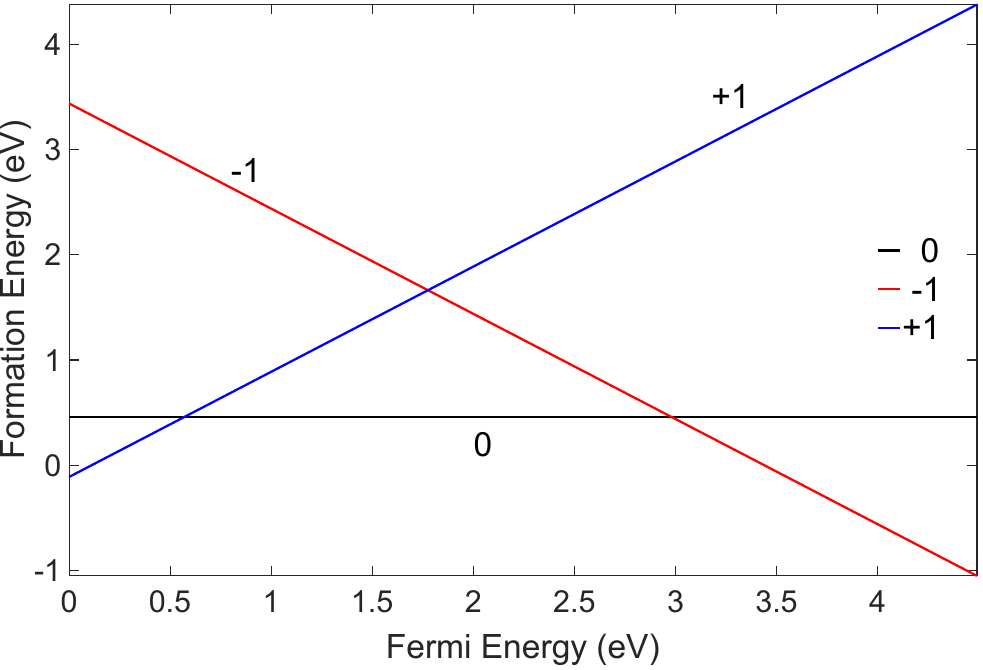}
	\caption{\label{FE-SiB} {\bf Formation energy of Si$_\mathrm{B}$.} Formation energy as a function of the Fermi energy for different charge states in N-rich environment.}
\end{figure}

The calculated Si$_\mathrm{B}$ formation energies in N-rich condition are shown in Fig.~\ref{FE-SiB}. The formation energy for the neutral defect is around 0.46~eV, which is higher than the previously reported value of -0.29~eV, possibly due to the small vacuum size (15~\text{\AA}) used in the earlier calculations~\cite{Mapasha2016,Mapasha22016}. However, the most stable charge state for Si$_\textrm{B}$ substitution is +1 which undergoes a slightly exothermic process. The defect energy transition states for $e$(+1/0) and $e$(0/-1) are expected to be at 0.61~eV and 2.96~eV with respect to VBM as calculated with the Perdew-Burke-Ernzerdorf (PBE)~\cite{Perdew1996} functional. Taking into account the difference in the band gaps calculated with PBE and HSE, as described above, we would expect the actual transition levels to be 1.23~eV and 3.58~eV
above VBM. Both defect levels are considered deep and expected to influence the optical properties of h-BN.

For h-BN layer which is on top of Cu(111), we expect that the copper (as during growth) Fermi level lies around $1.4-1.5$~eV above the VBM of h-BN~\cite{Laskowski2008,Zhang2018} and therefore the Si$_\mathrm{B}$ should be in neutral state. However it has been shown that under electron beam, BN-nanotubes are positively charged, which is attributed to emission of secondary electrons that shift the Fermi level to VBM and the Si$_\mathrm{B}$ switches to +1 state \cite{Wei2013}.    

\begin{figure}[htbp]
	\includegraphics[width=.40\textwidth]{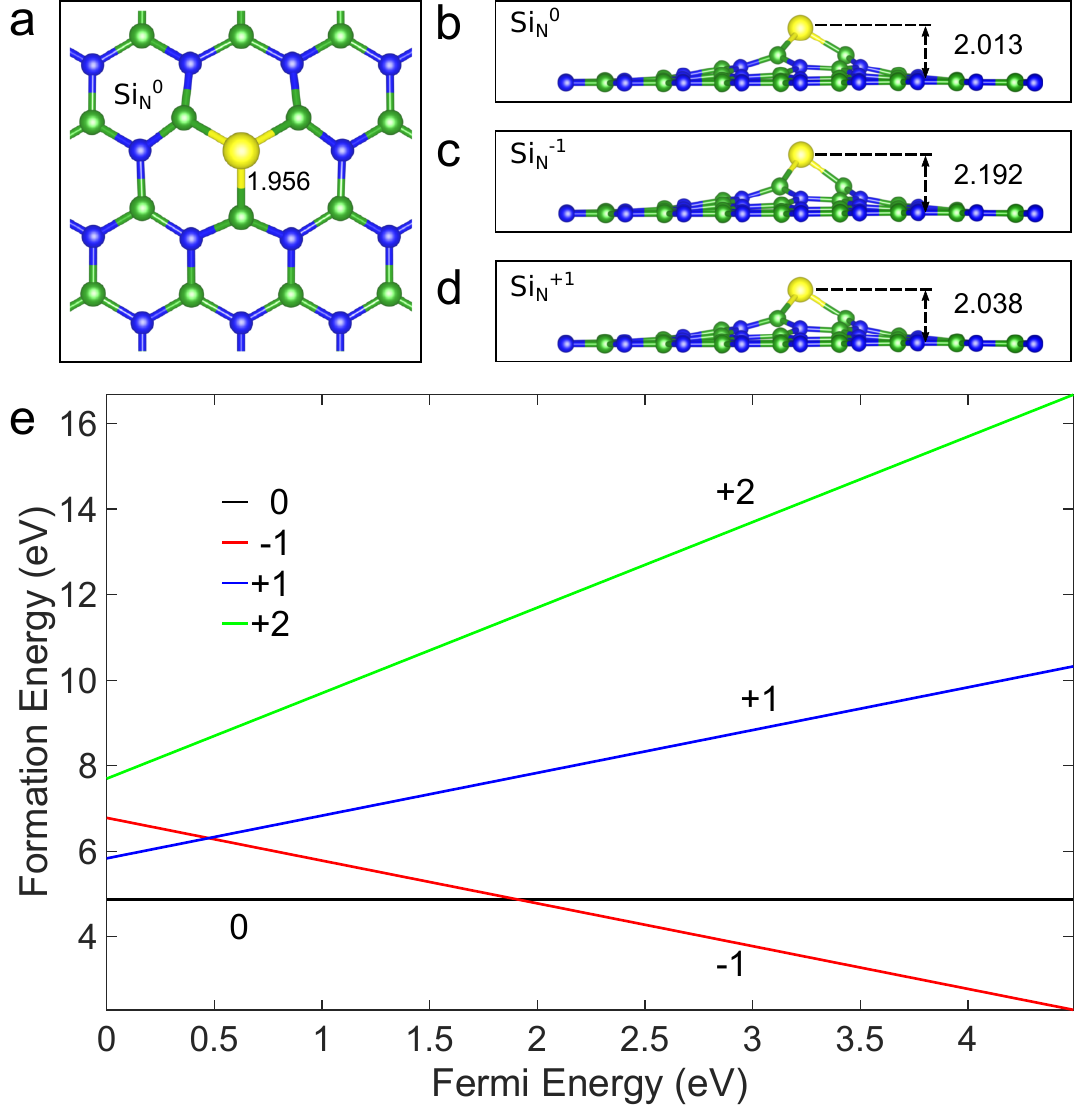}
	\caption{\label{SiN} {\bf Structure and formation energy of Si$_\mathrm{N}$.} (a) Top view of locally optimized Si$_\mathrm{N}^0$ substitution with Si-B bond length and (b) elevation of the silicon atom. (c,d) Side views of the charged structures. (e) Formation energy for Si$_\mathrm{N}$ in a boron rich environment.} 
\end{figure}

Fig.~\ref{SiN}a-b show the locally optimized structure of silicon substitution in nitrogen site in h-BN in its neutral state. Compared to Si$_\mathrm{B}$, in Si$_\mathrm{N}$ the silicon atom is extruded much higher out from h-BN. The distance between the silicon and the h-BN plane is $\sim{2.0}$ $\text{\AA}$ and the bond length between silicon and neighboring boron atom is 1.956~$\text{\AA}$. This buckling could be attributed to the electrostatic repulsion between silicon and neighboring boron atoms. Bader analysis~\cite{Tang2010} of charge density shows that the silicon atom has lost all valence electrons while the neighboring boron atoms have lost a fraction of their electrons to the neighboring nitrogen atoms. Adding an electron (Fig.~\ref{SiN}) would again slightly elevate the silicon atom. However, most of the charge is again transferred to the nitrogen atoms.  

The formation energy for Si$_\mathrm{N}$ in boron rich environment is shown in Fig.~\ref{SiN}e.  The formation energy for neutral defect is around 4.86~eV, which high compared to silicon in boron vacancy. Based on band structure calculation (Supplementary Material), possible charge states range from -1 to +2. The most stable case is when the silicon atom is in neutral state. The transition level $e$(0/-1) is at 1.90~eV as calculated with PBE. By rescaling the Fermi energy based on the calculated HSE band gap, we expected that this transition energy would rise to 2.51~eV.

\begin{figure}[htbp]
	\includegraphics[width=.40\textwidth]{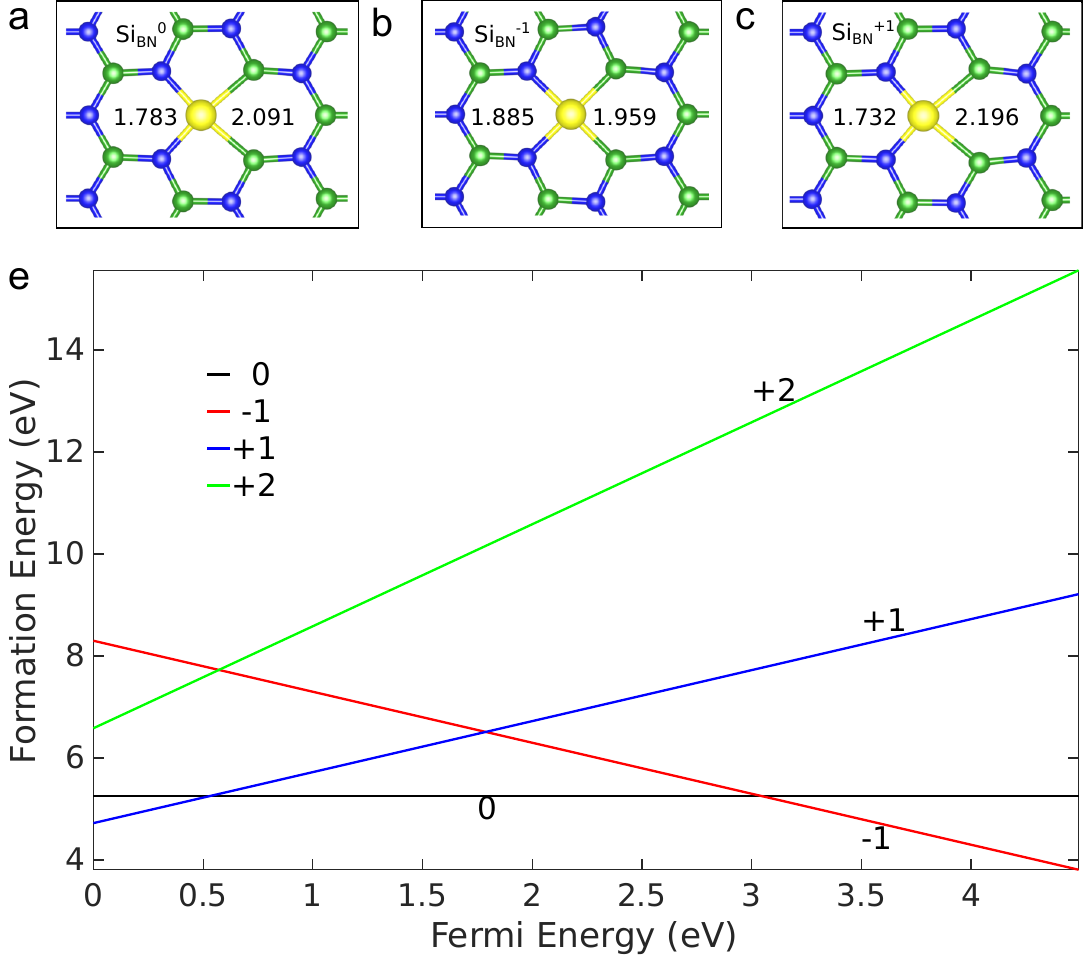}
	\caption{\label{SiDV}{\bf Structure and formation energy of Si$_\mathrm{BN}$.} Optimized model for silicon in (a) neutral (Si$_\mathrm{BN}^0$), (b) negatively charged (Si$_\mathrm{BN}^{-1}$) and (c) positively charged (Si$_\mathrm{BN}^{+1}$) states in a double vacancy. (e) Formation energy of silicon in a double vacancy.}
\end{figure}

We also calculated formation energy for silicon substitution in a double vacancy, where adjacent boron and nitrogen atoms are missing. The optimized structures are shown in Fig.~\ref{SiDV}. In this case the h-BN monolayer stays almost flat with the silicon-boron bond length of $\sim{2.09}$~$\text{\AA}$ and silicon-nitrogen bond length of $\sim{1.78}$~$\text{\AA}$.
One added electron changes the bond length for Si-B and Si-N to $\sim{1.96}$~$\text{\AA}$ and $\sim{1.88}$~$\text{\AA}$, respectively. From the band structure calculation (Supplementary Material) it is evident that both +1 and +2 states are possible. However, the silicon is pushed even more toward the nitrogen atoms due to electrostatic repulsion between boron atoms and the impurity. The bond lengths in this case are $\sim{2.20}$~$\text{\AA}$ and $\sim{1.73}$~$\text{\AA}$ for Si-B and Si-N bonds, respectively.

The calculated formation energies are shown in Fig.~\ref{SiDV}e. Interestingly, for the neutral structure of Si$_\mathrm{{BN}}$, the formation energy is around 5.25 eV, which is slightly higher than Si$_\mathrm{N}$ substitution despite the flat structure. The +2 charge state has the highest formation energy, and the PBE calculated transition levels are 0.51 eV for (+1/0) and 3.11 eV with respect to VBM. For HSE method of calculation, we would expect transition levels for $e$(+1/0) and $e$(0/-1)  to be 1.12~eV form VBM and 2.00~eV
from CBM.

We further calculated the migration barrier for Si$_\mathrm{B}$ to pass from one side of the h-BN plane to the other, a process that has been recently observed in graphene~\cite{hofer_direct_2019}. The barrier was calcalated with the nudged elastic band method~\cite{Henkelman2000}. The estimated energy barrier of 2.42 eV is much higher than 1.08 eV for silicon substitution in graphene~\cite{Susi2014}, which could be attributed to the ionic bonding between silicon and h-BN. The same transition can be achieved by providing the Si atom 3.28~eV of kinetic energy, as confirmed with molecular dynamics calculation with DFT (0.5~fs time step). This amount of energy can be transferred to a static Si atom by an electron with a kinetic energy of 40.5~keV and easily achieved under our experimental conditions (acceleration voltage 60~kV). Thus most Si atoms are expected to face away from the electron beam.

In this work we present the direct experimental observation of Si impurities in a free standing monolayer of hexagonal boron nitride (h-BN) using scanning transmission electron microscopy. Our density functional theory calculations show that Si atoms have the lowest formation energy in a boron vacancy, which is also the only experimentally observed configuration. Our simulations indicate +1 as the most likely charge state. Although the origin of Si atoms in our samples remains unknown, the results expands the number of observed impurity atoms in h-BN from the previously reported C and O to an element from the third row of the periodic table. This shows that heteroatom doping of h-BN with heavier elements is possible similar to graphene~\cite{Susi2017,Tripathi2018}, opening the way towards applications ranging from single-atom catalysis to atomically small magnetic structures.

We acknowledge the Austrian Science Fund (FWF) for funding through projects I3181 and P31605 as well as the Vienna Science and Technology Fund through project MA14-009. HPK acknowledges Academy of Finland for the support under Project No. 311058. We further acknowledge the Vienna Scientific Cluster for generous grants of computational resources. References\cite{krivanek_electron_2008,leuthner_scanning_2018,Kresse1993,Perdew1996,Kresse1999,komsa2014, Komsa2018,Heyd2003,Berseneva2011,Aliaksandr2006,Koch2002,Henkelman2000} are cited in the supplement.

\bibliography{references}

\begin{thebibliography}{51}%
\makeatletter
\providecommand \@ifxundefined [1]{%
 \@ifx{#1\undefined}
}%
\providecommand \@ifnum [1]{%
 \ifnum #1\expandafter \@firstoftwo
 \else \expandafter \@secondoftwo
 \fi
}%
\providecommand \@ifx [1]{%
 \ifx #1\expandafter \@firstoftwo
 \else \expandafter \@secondoftwo
 \fi
}%
\providecommand \natexlab [1]{#1}%
\providecommand \enquote  [1]{``#1''}%
\providecommand \bibnamefont  [1]{#1}%
\providecommand \bibfnamefont [1]{#1}%
\providecommand \citenamefont [1]{#1}%
\providecommand \href@noop [0]{\@secondoftwo}%
\providecommand \href [0]{\begingroup \@sanitize@url \@href}%
\providecommand \@href[1]{\@@startlink{#1}\@@href}%
\providecommand \@@href[1]{\endgroup#1\@@endlink}%
\providecommand \@sanitize@url [0]{\catcode `\\12\catcode `\$12\catcode
  `\&12\catcode `\#12\catcode `\^12\catcode `\_12\catcode `\%12\relax}%
\providecommand \@@startlink[1]{}%
\providecommand \@@endlink[0]{}%
\providecommand \url  [0]{\begingroup\@sanitize@url \@url }%
\providecommand \@url [1]{\endgroup\@href {#1}{\urlprefix }}%
\providecommand \urlprefix  [0]{URL }%
\providecommand \Eprint [0]{\href }%
\providecommand \doibase [0]{http://dx.doi.org/}%
\providecommand \selectlanguage [0]{\@gobble}%
\providecommand \bibinfo  [0]{\@secondoftwo}%
\providecommand \bibfield  [0]{\@secondoftwo}%
\providecommand \translation [1]{[#1]}%
\providecommand \BibitemOpen [0]{}%
\providecommand \bibitemStop [0]{}%
\providecommand \bibitemNoStop [0]{.\EOS\space}%
\providecommand \EOS [0]{\spacefactor3000\relax}%
\providecommand \BibitemShut  [1]{\csname bibitem#1\endcsname}%
\let\auto@bib@innerbib\@empty
\bibitem [{\citenamefont {Novoselov}\ \emph {et~al.}(2004)\citenamefont
  {Novoselov}, \citenamefont {Geim}, \citenamefont {Morozov}, \citenamefont
  {Jiang}, \citenamefont {Zhang}, \citenamefont {Dubonos}, \citenamefont
  {Grigorieva},\ and\ \citenamefont {Firsov}}]{Novoselov2004}%
  \BibitemOpen
  \bibfield  {author} {\bibinfo {author} {\bibfnamefont {K.~S.}\ \bibnamefont
  {Novoselov}}, \bibinfo {author} {\bibfnamefont {A.~K.}\ \bibnamefont {Geim}},
  \bibinfo {author} {\bibfnamefont {S.~V.}\ \bibnamefont {Morozov}}, \bibinfo
  {author} {\bibfnamefont {D.}~\bibnamefont {Jiang}}, \bibinfo {author}
  {\bibfnamefont {Y.}~\bibnamefont {Zhang}}, \bibinfo {author} {\bibfnamefont
  {S.~V.}\ \bibnamefont {Dubonos}}, \bibinfo {author} {\bibfnamefont {I.~V.}\
  \bibnamefont {Grigorieva}}, \ and\ \bibinfo {author} {\bibfnamefont {A.~A.}\
  \bibnamefont {Firsov}},\ }\bibfield  {title} {\enquote {\bibinfo {title}
  {Electric field effect in atomically thin carbon films},}\ }\href {\doibase
  10.1126/science.1102896} {\bibfield  {journal} {\bibinfo  {journal}
  {Science}\ }\textbf {\bibinfo {volume} {306}},\ \bibinfo {pages} {666--669}
  (\bibinfo {year} {2004})}\BibitemShut {NoStop}%
\bibitem [{\citenamefont {Pacil\'{e}}\ \emph {et~al.}(2008)\citenamefont
  {Pacil\'{e}}, \citenamefont {Meyer}, \citenamefont {Girit},\ and\
  \citenamefont {Zettl}}]{Pacile2008}%
  \BibitemOpen
  \bibfield  {author} {\bibinfo {author} {\bibfnamefont {D.}~\bibnamefont
  {Pacil\'{e}}}, \bibinfo {author} {\bibfnamefont {J.~C.}\ \bibnamefont
  {Meyer}}, \bibinfo {author} {\bibfnamefont {c.~O.}\ \bibnamefont {Girit}}, \
  and\ \bibinfo {author} {\bibfnamefont {A.}~\bibnamefont {Zettl}},\ }\bibfield
   {title} {\enquote {\bibinfo {title} {The two-dimensional phase of boron
  nitride: Few-atomic-layer sheets and suspended membranes},}\ }\href {\doibase
  10.1063/1.2903702} {\bibfield  {journal} {\bibinfo  {journal} {Applied
  Physics Letters}\ }\textbf {\bibinfo {volume} {92}},\ \bibinfo {pages}
  {133107} (\bibinfo {year} {2008})}\BibitemShut {NoStop}%
\bibitem [{\citenamefont {Wang}\ \emph {et~al.}(2012)\citenamefont {Wang},
  \citenamefont {Kalantar-Zadeh}, \citenamefont {Kis}, \citenamefont
  {Coleman},\ and\ \citenamefont {Strano}}]{Wang2012}%
  \BibitemOpen
  \bibfield  {author} {\bibinfo {author} {\bibfnamefont {Q.~H.}\ \bibnamefont
  {Wang}}, \bibinfo {author} {\bibfnamefont {K.}~\bibnamefont
  {Kalantar-Zadeh}}, \bibinfo {author} {\bibfnamefont {A.}~\bibnamefont {Kis}},
  \bibinfo {author} {\bibfnamefont {J.~N.}\ \bibnamefont {Coleman}}, \ and\
  \bibinfo {author} {\bibfnamefont {M.~S.}\ \bibnamefont {Strano}},\ }\bibfield
   {title} {\enquote {\bibinfo {title} {{Electronics and optoelectronics of
  two-dimensional transition metal dichalcogenides}},}\ }\href {\doibase
  10.1038/nnano.2012.193} {\bibfield  {journal} {\bibinfo  {journal} {Nature
  Nanotechnology}\ }\textbf {\bibinfo {volume} {7}},\ \bibinfo {pages}
  {699--712} (\bibinfo {year} {2012})}\BibitemShut {NoStop}%
\bibitem [{\citenamefont {Cassabois}, \citenamefont {Valvin},\ and\
  \citenamefont {Gil}(2016)}]{Cassabois2016}%
  \BibitemOpen
  \bibfield  {author} {\bibinfo {author} {\bibfnamefont {G.}~\bibnamefont
  {Cassabois}}, \bibinfo {author} {\bibfnamefont {P.}~\bibnamefont {Valvin}}, \
  and\ \bibinfo {author} {\bibfnamefont {B.}~\bibnamefont {Gil}},\ }\bibfield
  {title} {\enquote {\bibinfo {title} {{Hexagonal boron nitride is an indirect
  bandgap semiconductor}},}\ }\href {\doibase 10.1038/nphoton.2015.277}
  {\bibfield  {journal} {\bibinfo  {journal} {Nature Photonics}\ }\textbf
  {\bibinfo {volume} {10}},\ \bibinfo {pages} {262--266} (\bibinfo {year}
  {2016})}\BibitemShut {NoStop}%
\bibitem [{\citenamefont {Geim}\ and\ \citenamefont
  {Grigorieva}(2013)}]{Geim2013}%
  \BibitemOpen
  \bibfield  {author} {\bibinfo {author} {\bibfnamefont {A.~K.}\ \bibnamefont
  {Geim}}\ and\ \bibinfo {author} {\bibfnamefont {I.~V.}\ \bibnamefont
  {Grigorieva}},\ }\bibfield  {title} {\enquote {\bibinfo {title} {{Van der
  Waals heterostructures}},}\ }\href {\doibase 10.1038/nature12385} {\bibfield
  {journal} {\bibinfo  {journal} {Nature}\ }\textbf {\bibinfo {volume} {499}},\
  \bibinfo {pages} {419--425} (\bibinfo {year} {2013})}\BibitemShut {NoStop}%
\bibitem [{\citenamefont {Oba}\ and\ \citenamefont {Kumagai}(2018)}]{Oba2018}%
  \BibitemOpen
  \bibfield  {author} {\bibinfo {author} {\bibfnamefont {F.}~\bibnamefont
  {Oba}}\ and\ \bibinfo {author} {\bibfnamefont {Y.}~\bibnamefont {Kumagai}},\
  }\bibfield  {title} {\enquote {\bibinfo {title} {Design and exploration of
  semiconductors from first principles: A review of recent advances},}\ }\href
  {http://stacks.iop.org/1882-0786/11/i=6/a=060101} {\bibfield  {journal}
  {\bibinfo  {journal} {Applied Physics Express}\ }\textbf {\bibinfo {volume}
  {11}},\ \bibinfo {pages} {060101} (\bibinfo {year} {2018})}\BibitemShut
  {NoStop}%
\bibitem [{\citenamefont {Kapetanakis}\ \emph {et~al.}(2016)\citenamefont
  {Kapetanakis}, \citenamefont {Oxley}, \citenamefont {Zhou}, \citenamefont
  {Pennycook}, \citenamefont {Idrobo},\ and\ \citenamefont
  {Pantelides}}]{Kapetanakis2016}%
  \BibitemOpen
  \bibfield  {author} {\bibinfo {author} {\bibfnamefont {M.~D.}\ \bibnamefont
  {Kapetanakis}}, \bibinfo {author} {\bibfnamefont {M.~P.}\ \bibnamefont
  {Oxley}}, \bibinfo {author} {\bibfnamefont {W.}~\bibnamefont {Zhou}},
  \bibinfo {author} {\bibfnamefont {S.~J.}\ \bibnamefont {Pennycook}}, \bibinfo
  {author} {\bibfnamefont {J.-C.}\ \bibnamefont {Idrobo}}, \ and\ \bibinfo
  {author} {\bibfnamefont {S.~T.}\ \bibnamefont {Pantelides}},\ }\bibfield
  {title} {\enquote {\bibinfo {title} {Signatures of distinct impurity
  configurations in atomic-resolution valence electron-energy-loss
  spectroscopy: Application to graphene},}\ }\href {\doibase
  10.1103/PhysRevB.94.155449} {\bibfield  {journal} {\bibinfo  {journal} {Phys.
  Rev. B}\ }\textbf {\bibinfo {volume} {94}},\ \bibinfo {pages} {155449}
  (\bibinfo {year} {2016})}\BibitemShut {NoStop}%
\bibitem [{\citenamefont {Kotakoski}\ \emph {et~al.}(2011)\citenamefont
  {Kotakoski}, \citenamefont {Krasheninnikov}, \citenamefont {Kaiser},\ and\
  \citenamefont {Meyer}}]{Kotakoski2011}%
  \BibitemOpen
  \bibfield  {author} {\bibinfo {author} {\bibfnamefont {J.}~\bibnamefont
  {Kotakoski}}, \bibinfo {author} {\bibfnamefont {A.~V.}\ \bibnamefont
  {Krasheninnikov}}, \bibinfo {author} {\bibfnamefont {U.}~\bibnamefont
  {Kaiser}}, \ and\ \bibinfo {author} {\bibfnamefont {J.~C.}\ \bibnamefont
  {Meyer}},\ }\bibfield  {title} {\enquote {\bibinfo {title} {From point
  defects in graphene to two-dimensional amorphous carbon},}\ }\href {\doibase
  10.1103/PhysRevLett.106.105505} {\bibfield  {journal} {\bibinfo  {journal}
  {Phys. Rev. Lett.}\ }\textbf {\bibinfo {volume} {106}},\ \bibinfo {pages}
  {105505} (\bibinfo {year} {2011})}\BibitemShut {NoStop}%
\bibitem [{\citenamefont {Hu}\ \emph {et~al.}(2018)\citenamefont {Hu},
  \citenamefont {Wu}, \citenamefont {Han}, \citenamefont {He}, \citenamefont
  {Ni},\ and\ \citenamefont {Chen}}]{Hu2018}%
  \BibitemOpen
  \bibfield  {author} {\bibinfo {author} {\bibfnamefont {Z.}~\bibnamefont
  {Hu}}, \bibinfo {author} {\bibfnamefont {Z.}~\bibnamefont {Wu}}, \bibinfo
  {author} {\bibfnamefont {C.}~\bibnamefont {Han}}, \bibinfo {author}
  {\bibfnamefont {J.}~\bibnamefont {He}}, \bibinfo {author} {\bibfnamefont
  {Z.}~\bibnamefont {Ni}}, \ and\ \bibinfo {author} {\bibfnamefont
  {W.}~\bibnamefont {Chen}},\ }\bibfield  {title} {\enquote {\bibinfo {title}
  {Two-dimensional transition metal dichalcogenides: interface and defect
  engineering},}\ }\href {\doibase 10.1039/C8CS00024G} {\bibfield  {journal}
  {\bibinfo  {journal} {Chem. Soc. Rev.}\ }\textbf {\bibinfo {volume} {47}},\
  \bibinfo {pages} {3100--3128} (\bibinfo {year} {2018})}\BibitemShut {NoStop}%
\bibitem [{\citenamefont {Zhou}\ \emph {et~al.}(2012)\citenamefont {Zhou},
  \citenamefont {Kapetanakis}, \citenamefont {Prange}, \citenamefont
  {Pantelides}, \citenamefont {Pennycook},\ and\ \citenamefont
  {Idrobo}}]{Zhou2012}%
  \BibitemOpen
  \bibfield  {author} {\bibinfo {author} {\bibfnamefont {W.}~\bibnamefont
  {Zhou}}, \bibinfo {author} {\bibfnamefont {M.~D.}\ \bibnamefont
  {Kapetanakis}}, \bibinfo {author} {\bibfnamefont {M.~P.}\ \bibnamefont
  {Prange}}, \bibinfo {author} {\bibfnamefont {S.~T.}\ \bibnamefont
  {Pantelides}}, \bibinfo {author} {\bibfnamefont {S.~J.}\ \bibnamefont
  {Pennycook}}, \ and\ \bibinfo {author} {\bibfnamefont {J.-C.}\ \bibnamefont
  {Idrobo}},\ }\bibfield  {title} {\enquote {\bibinfo {title} {Direct
  determination of the chemical bonding of individual impurities in
  graphene},}\ }\href {\doibase 10.1103/PhysRevLett.109.206803} {\bibfield
  {journal} {\bibinfo  {journal} {Phys. Rev. Lett.}\ }\textbf {\bibinfo
  {volume} {109}},\ \bibinfo {pages} {206803} (\bibinfo {year}
  {2012})}\BibitemShut {NoStop}%
\bibitem [{\citenamefont {Ramasse}\ \emph
  {et~al.}(2013{\natexlab{a}})\citenamefont {Ramasse}, \citenamefont
  {Seabourne}, \citenamefont {Kepaptsoglou}, \citenamefont {Zan}, \citenamefont
  {Bangert},\ and\ \citenamefont {Scott}}]{Ramasse2013}%
  \BibitemOpen
  \bibfield  {author} {\bibinfo {author} {\bibfnamefont {Q.~M.}\ \bibnamefont
  {Ramasse}}, \bibinfo {author} {\bibfnamefont {C.~R.}\ \bibnamefont
  {Seabourne}}, \bibinfo {author} {\bibfnamefont {D.-M.}\ \bibnamefont
  {Kepaptsoglou}}, \bibinfo {author} {\bibfnamefont {R.}~\bibnamefont {Zan}},
  \bibinfo {author} {\bibfnamefont {U.}~\bibnamefont {Bangert}}, \ and\
  \bibinfo {author} {\bibfnamefont {A.~J.}\ \bibnamefont {Scott}},\ }\bibfield
  {title} {\enquote {\bibinfo {title} {{Probing the Bonding and Electronic
  Structure of Single Atom Dopants in Graphene with Electron Energy Loss
  Spectroscopy}},}\ }\href {\doibase 10.1021/nl304187e} {\bibfield  {journal}
  {\bibinfo  {journal} {Nano Letters}\ }\textbf {\bibinfo {volume} {13}},\
  \bibinfo {pages} {4989--4995} (\bibinfo {year}
  {2013}{\natexlab{a}})}\BibitemShut {NoStop}%
\bibitem [{\citenamefont {Jalili}\ \emph {et~al.}(2018)\citenamefont {Jalili},
  \citenamefont {Esrafilzadeh}, \citenamefont {Aboutalebi}, \citenamefont
  {Sabri}, \citenamefont {Kandjani}, \citenamefont {Bhargava}, \citenamefont
  {{Della Gaspera}}, \citenamefont {Gengenbach}, \citenamefont {Walker},
  \citenamefont {Chao}, \citenamefont {Wang}, \citenamefont {Alimadadi},
  \citenamefont {Mitchell}, \citenamefont {Officer}, \citenamefont
  {MacFarlane},\ and\ \citenamefont {Wallace}}]{Jalili2018}%
  \BibitemOpen
  \bibfield  {author} {\bibinfo {author} {\bibfnamefont {R.}~\bibnamefont
  {Jalili}}, \bibinfo {author} {\bibfnamefont {D.}~\bibnamefont
  {Esrafilzadeh}}, \bibinfo {author} {\bibfnamefont {S.~H.}\ \bibnamefont
  {Aboutalebi}}, \bibinfo {author} {\bibfnamefont {Y.~M.}\ \bibnamefont
  {Sabri}}, \bibinfo {author} {\bibfnamefont {A.~E.}\ \bibnamefont {Kandjani}},
  \bibinfo {author} {\bibfnamefont {S.~K.}\ \bibnamefont {Bhargava}}, \bibinfo
  {author} {\bibfnamefont {E.}~\bibnamefont {{Della Gaspera}}}, \bibinfo
  {author} {\bibfnamefont {T.~R.}\ \bibnamefont {Gengenbach}}, \bibinfo
  {author} {\bibfnamefont {A.}~\bibnamefont {Walker}}, \bibinfo {author}
  {\bibfnamefont {Y.}~\bibnamefont {Chao}}, \bibinfo {author} {\bibfnamefont
  {C.}~\bibnamefont {Wang}}, \bibinfo {author} {\bibfnamefont {H.}~\bibnamefont
  {Alimadadi}}, \bibinfo {author} {\bibfnamefont {D.~R.~G.}\ \bibnamefont
  {Mitchell}}, \bibinfo {author} {\bibfnamefont {D.~L.}\ \bibnamefont
  {Officer}}, \bibinfo {author} {\bibfnamefont {D.~R.}\ \bibnamefont
  {MacFarlane}}, \ and\ \bibinfo {author} {\bibfnamefont {G.~G.}\ \bibnamefont
  {Wallace}},\ }\bibfield  {title} {\enquote {\bibinfo {title} {{Silicon as a
  ubiquitous contaminant in graphene derivatives with significant impact on
  device performance}},}\ }\href {\doibase 10.1038/s41467-018-07396-3}
  {\bibfield  {journal} {\bibinfo  {journal} {Nature Communications}\ }\textbf
  {\bibinfo {volume} {9}},\ \bibinfo {pages} {5070} (\bibinfo {year}
  {2018})}\BibitemShut {NoStop}%
\bibitem [{\citenamefont {Krivanek}\ \emph {et~al.}(2010)\citenamefont
  {Krivanek}, \citenamefont {Chisholm}, \citenamefont {Nicolosi}, \citenamefont
  {Pennycook}, \citenamefont {Corbin}, \citenamefont {Dellby}, \citenamefont
  {Murfitt}, \citenamefont {Own}, \citenamefont {Szilagyi}, \citenamefont
  {Oxley}, \citenamefont {Pantelides},\ and\ \citenamefont
  {Pennycook}}]{Krivanek2010}%
  \BibitemOpen
  \bibfield  {author} {\bibinfo {author} {\bibfnamefont {O.~L.}\ \bibnamefont
  {Krivanek}}, \bibinfo {author} {\bibfnamefont {M.~F.}\ \bibnamefont
  {Chisholm}}, \bibinfo {author} {\bibfnamefont {V.}~\bibnamefont {Nicolosi}},
  \bibinfo {author} {\bibfnamefont {T.~J.}\ \bibnamefont {Pennycook}}, \bibinfo
  {author} {\bibfnamefont {G.~J.}\ \bibnamefont {Corbin}}, \bibinfo {author}
  {\bibfnamefont {N.}~\bibnamefont {Dellby}}, \bibinfo {author} {\bibfnamefont
  {M.~F.}\ \bibnamefont {Murfitt}}, \bibinfo {author} {\bibfnamefont {C.~S.}\
  \bibnamefont {Own}}, \bibinfo {author} {\bibfnamefont {Z.~S.}\ \bibnamefont
  {Szilagyi}}, \bibinfo {author} {\bibfnamefont {M.~P.}\ \bibnamefont {Oxley}},
  \bibinfo {author} {\bibfnamefont {S.~T.}\ \bibnamefont {Pantelides}}, \ and\
  \bibinfo {author} {\bibfnamefont {S.~J.}\ \bibnamefont {Pennycook}},\
  }\bibfield  {title} {\enquote {\bibinfo {title} {{Atom-by-atom structural and
  chemical analysis by annular dark-field electron microscopy}},}\ }\href
  {\doibase 10.1038/nature08879} {\bibfield  {journal} {\bibinfo  {journal}
  {Nature}\ }\textbf {\bibinfo {volume} {464}},\ \bibinfo {pages} {571--574}
  (\bibinfo {year} {2010})}\BibitemShut {NoStop}%
\bibitem [{\citenamefont {Meyer}\ \emph {et~al.}(2009)\citenamefont {Meyer},
  \citenamefont {Chuvilin}, \citenamefont {Algara-Siller}, \citenamefont
  {Biskupek},\ and\ \citenamefont {Kaiser}}]{Meyer2009}%
  \BibitemOpen
  \bibfield  {author} {\bibinfo {author} {\bibfnamefont {J.~C.}\ \bibnamefont
  {Meyer}}, \bibinfo {author} {\bibfnamefont {A.}~\bibnamefont {Chuvilin}},
  \bibinfo {author} {\bibfnamefont {G.}~\bibnamefont {Algara-Siller}}, \bibinfo
  {author} {\bibfnamefont {J.}~\bibnamefont {Biskupek}}, \ and\ \bibinfo
  {author} {\bibfnamefont {U.}~\bibnamefont {Kaiser}},\ }\bibfield  {title}
  {\enquote {\bibinfo {title} {Selective sputtering and atomic resolution
  imaging of atomically thin boron nitride membranes},}\ }\href {\doibase
  10.1021/nl9011497} {\bibfield  {journal} {\bibinfo  {journal} {Nano Letters}\
  }\textbf {\bibinfo {volume} {9}},\ \bibinfo {pages} {2683--2689} (\bibinfo
  {year} {2009})}\BibitemShut {NoStop}%
\bibitem [{\citenamefont {Kim}\ \emph {et~al.}(2018)\citenamefont {Kim},
  \citenamefont {Jeong}, \citenamefont {Kim}, \citenamefont {Han},\ and\
  \citenamefont {Kim}}]{Kim2018}%
  \BibitemOpen
  \bibfield  {author} {\bibinfo {author} {\bibfnamefont {D.~Y.}\ \bibnamefont
  {Kim}}, \bibinfo {author} {\bibfnamefont {H.}~\bibnamefont {Jeong}}, \bibinfo
  {author} {\bibfnamefont {J.}~\bibnamefont {Kim}}, \bibinfo {author}
  {\bibfnamefont {N.}~\bibnamefont {Han}}, \ and\ \bibinfo {author}
  {\bibfnamefont {J.~K.}\ \bibnamefont {Kim}},\ }\bibfield  {title} {\enquote
  {\bibinfo {title} {Defect-mediated in-plane electrical conduction in
  few-layer sp2-hybridized boron nitrides},}\ }\href {\doibase
  10.1021/acsami.8b04389} {\bibfield  {journal} {\bibinfo  {journal} {ACS
  Applied Materials \& Interfaces}\ }\textbf {\bibinfo {volume} {10}},\
  \bibinfo {pages} {17287--17294} (\bibinfo {year} {2018})}\BibitemShut
  {NoStop}%
\bibitem [{\citenamefont {Murata}\ \emph {et~al.}(2013)\citenamefont {Murata},
  \citenamefont {Taniguchi}, \citenamefont {Hishita}, \citenamefont {Yamamoto},
  \citenamefont {Oba},\ and\ \citenamefont {Tanaka}}]{Murata2013}%
  \BibitemOpen
  \bibfield  {author} {\bibinfo {author} {\bibfnamefont {H.}~\bibnamefont
  {Murata}}, \bibinfo {author} {\bibfnamefont {T.}~\bibnamefont {Taniguchi}},
  \bibinfo {author} {\bibfnamefont {S.}~\bibnamefont {Hishita}}, \bibinfo
  {author} {\bibfnamefont {T.}~\bibnamefont {Yamamoto}}, \bibinfo {author}
  {\bibfnamefont {F.}~\bibnamefont {Oba}}, \ and\ \bibinfo {author}
  {\bibfnamefont {I.}~\bibnamefont {Tanaka}},\ }\bibfield  {title} {\enquote
  {\bibinfo {title} {Local environment of silicon in cubic boron nitride},}\
  }\href@noop {} {\bibfield  {journal} {\bibinfo  {journal} {Journal of Applied
  Physics}\ }\textbf {\bibinfo {volume} {114}} (\bibinfo {year}
  {2013})}\BibitemShut {NoStop}%
\bibitem [{\citenamefont {Majety}\ \emph {et~al.}(2013)\citenamefont {Majety},
  \citenamefont {Doan}, \citenamefont {Li}, \citenamefont {Lin},\ and\
  \citenamefont {Jiang}}]{Majety2013}%
  \BibitemOpen
  \bibfield  {author} {\bibinfo {author} {\bibfnamefont {S.}~\bibnamefont
  {Majety}}, \bibinfo {author} {\bibfnamefont {T.~C.}\ \bibnamefont {Doan}},
  \bibinfo {author} {\bibfnamefont {J.}~\bibnamefont {Li}}, \bibinfo {author}
  {\bibfnamefont {J.~Y.}\ \bibnamefont {Lin}}, \ and\ \bibinfo {author}
  {\bibfnamefont {H.~X.}\ \bibnamefont {Jiang}},\ }\bibfield  {title} {\enquote
  {\bibinfo {title} {Electrical transport properties of si-doped hexagonal
  boron nitride epilayers},}\ }\href {\doibase 10.1063/1.4860949} {\bibfield
  {journal} {\bibinfo  {journal} {AIP Advances}\ }\textbf {\bibinfo {volume}
  {3}},\ \bibinfo {pages} {122116} (\bibinfo {year} {2013})}\BibitemShut
  {NoStop}%
\bibitem [{\citenamefont {Sajid}, \citenamefont {Reimers},\ and\ \citenamefont
  {Ford}(2018)}]{Sajid2018}%
  \BibitemOpen
  \bibfield  {author} {\bibinfo {author} {\bibfnamefont {A.}~\bibnamefont
  {Sajid}}, \bibinfo {author} {\bibfnamefont {J.~R.}\ \bibnamefont {Reimers}},
  \ and\ \bibinfo {author} {\bibfnamefont {M.~J.}\ \bibnamefont {Ford}},\
  }\bibfield  {title} {\enquote {\bibinfo {title} {Defect states in hexagonal
  boron nitride: Assignments of observed properties and prediction of
  properties relevant to quantum computation},}\ }\href {\doibase
  10.1103/PhysRevB.97.064101} {\bibfield  {journal} {\bibinfo  {journal} {Phys.
  Rev. B}\ }\textbf {\bibinfo {volume} {97}},\ \bibinfo {pages} {064101}
  (\bibinfo {year} {2018})}\BibitemShut {NoStop}%
\bibitem [{\citenamefont {Tang}\ and\ \citenamefont {Cao}(2010)}]{Tang2010}%
  \BibitemOpen
  \bibfield  {author} {\bibinfo {author} {\bibfnamefont {S.}~\bibnamefont
  {Tang}}\ and\ \bibinfo {author} {\bibfnamefont {Z.}~\bibnamefont {Cao}},\
  }\bibfield  {title} {\enquote {\bibinfo {title} {Carbon-doped zigzag boron
  nitride nanoribbons with widely tunable electronic and magnetic properties:
  insight from density functional calculations},}\ }\href {\doibase
  10.1039/B920754F} {\bibfield  {journal} {\bibinfo  {journal} {Phys. Chem.
  Chem. Phys.}\ }\textbf {\bibinfo {volume} {12}},\ \bibinfo {pages}
  {2313--2320} (\bibinfo {year} {2010})}\BibitemShut {NoStop}%
\bibitem [{\citenamefont {Asshoff}\ \emph {et~al.}(2018)\citenamefont
  {Asshoff}, \citenamefont {Sambricio}, \citenamefont {Slizovskiy},
  \citenamefont {Rooney}, \citenamefont {Taniguchi}, \citenamefont {Watanabe},
  \citenamefont {Haigh}, \citenamefont {Fal’ko}, \citenamefont {Grigorieva},\
  and\ \citenamefont {Vera-Marun}}]{Asshoff2018}%
  \BibitemOpen
  \bibfield  {author} {\bibinfo {author} {\bibfnamefont {P.~U.}\ \bibnamefont
  {Asshoff}}, \bibinfo {author} {\bibfnamefont {J.~L.}\ \bibnamefont
  {Sambricio}}, \bibinfo {author} {\bibfnamefont {S.}~\bibnamefont
  {Slizovskiy}}, \bibinfo {author} {\bibfnamefont {A.~P.}\ \bibnamefont
  {Rooney}}, \bibinfo {author} {\bibfnamefont {T.}~\bibnamefont {Taniguchi}},
  \bibinfo {author} {\bibfnamefont {K.}~\bibnamefont {Watanabe}}, \bibinfo
  {author} {\bibfnamefont {S.~J.}\ \bibnamefont {Haigh}}, \bibinfo {author}
  {\bibfnamefont {V.}~\bibnamefont {Fal’ko}}, \bibinfo {author}
  {\bibfnamefont {I.~V.}\ \bibnamefont {Grigorieva}}, \ and\ \bibinfo {author}
  {\bibfnamefont {I.~J.}\ \bibnamefont {Vera-Marun}},\ }\bibfield  {title}
  {\enquote {\bibinfo {title} {Magnetoresistance in co-hbn-nife tunnel
  junctions enhanced by resonant tunneling through single defects in ultrathin
  hbn barriers},}\ }\href {\doibase 10.1021/acs.nanolett.8b02866} {\bibfield
  {journal} {\bibinfo  {journal} {Nano Letters}\ }\textbf {\bibinfo {volume}
  {18}},\ \bibinfo {pages} {6954--6960} (\bibinfo {year} {2018})}\BibitemShut
  {NoStop}%
\bibitem [{\citenamefont {Wang}\ \emph {et~al.}(2018)\citenamefont {Wang},
  \citenamefont {Tang}, \citenamefont {Ren}, \citenamefont {Wang},
  \citenamefont {Han},\ and\ \citenamefont {Dai}}]{Wang2018}%
  \BibitemOpen
  \bibfield  {author} {\bibinfo {author} {\bibfnamefont {M.}~\bibnamefont
  {Wang}}, \bibinfo {author} {\bibfnamefont {S.}~\bibnamefont {Tang}}, \bibinfo
  {author} {\bibfnamefont {J.}~\bibnamefont {Ren}}, \bibinfo {author}
  {\bibfnamefont {B.}~\bibnamefont {Wang}}, \bibinfo {author} {\bibfnamefont
  {Y.}~\bibnamefont {Han}}, \ and\ \bibinfo {author} {\bibfnamefont
  {Y.}~\bibnamefont {Dai}},\ }\bibfield  {title} {\enquote {\bibinfo {title}
  {Magnetism in boron nitride monolayer induced by cobalt or nickel doping},}\
  }\href {\doibase 10.1007/s10948-017-4353-5} {\bibfield  {journal} {\bibinfo
  {journal} {Journal of Superconductivity and Novel Magnetism}\ }\textbf
  {\bibinfo {volume} {31}},\ \bibinfo {pages} {1559--1565} (\bibinfo {year}
  {2018})}\BibitemShut {NoStop}%
\bibitem [{\citenamefont {Liu}\ \emph {et~al.}(2014)\citenamefont {Liu},
  \citenamefont {Gao}, \citenamefont {Xu}, \citenamefont {Wang},\ and\
  \citenamefont {Zhao}}]{Liu2014}%
  \BibitemOpen
  \bibfield  {author} {\bibinfo {author} {\bibfnamefont {Y.-j.}\ \bibnamefont
  {Liu}}, \bibinfo {author} {\bibfnamefont {B.}~\bibnamefont {Gao}}, \bibinfo
  {author} {\bibfnamefont {D.}~\bibnamefont {Xu}}, \bibinfo {author}
  {\bibfnamefont {H.-m.}\ \bibnamefont {Wang}}, \ and\ \bibinfo {author}
  {\bibfnamefont {J.-x.}\ \bibnamefont {Zhao}},\ }\bibfield  {title} {\enquote
  {\bibinfo {title} {{Theoretical study on Si-doped hexagonal boron nitride ( h
  -BN) sheet: Electronic, magnetic properties, and reactivity}},}\ }\href
  {\doibase 10.1016/j.physleta.2014.07.053} {\bibfield  {journal} {\bibinfo
  {journal} {Physics Letters A}\ }\textbf {\bibinfo {volume} {378}},\ \bibinfo
  {pages} {2989--2994} (\bibinfo {year} {2014})}\BibitemShut {NoStop}%
\bibitem [{\citenamefont {Mapasha}, \citenamefont {Igumbor},\ and\
  \citenamefont {Chetty}(2016)}]{Mapasha22016}%
  \BibitemOpen
  \bibfield  {author} {\bibinfo {author} {\bibfnamefont {R.~E.}\ \bibnamefont
  {Mapasha}}, \bibinfo {author} {\bibfnamefont {E.}~\bibnamefont {Igumbor}}, \
  and\ \bibinfo {author} {\bibfnamefont {N.}~\bibnamefont {Chetty}},\
  }\bibfield  {title} {\enquote {\bibinfo {title} {A hybrid density functional
  study of silicon and phosphorus doped hexagonal boron nitride monolayer},}\
  }\href {http://stacks.iop.org/1742-6596/759/i=1/a=012042} {\bibfield
  {journal} {\bibinfo  {journal} {Journal of Physics: Conference Series}\
  }\textbf {\bibinfo {volume} {759}},\ \bibinfo {pages} {012042} (\bibinfo
  {year} {2016})}\BibitemShut {NoStop}%
\bibitem [{\citenamefont {Mapasha}\ \emph {et~al.}(2016)\citenamefont
  {Mapasha}, \citenamefont {Molepo}, \citenamefont {Andrew},\ and\
  \citenamefont {Chetty}}]{Mapasha2016}%
  \BibitemOpen
  \bibfield  {author} {\bibinfo {author} {\bibfnamefont {R.~E.}\ \bibnamefont
  {Mapasha}}, \bibinfo {author} {\bibfnamefont {M.~P.}\ \bibnamefont {Molepo}},
  \bibinfo {author} {\bibfnamefont {R.~C.}\ \bibnamefont {Andrew}}, \ and\
  \bibinfo {author} {\bibfnamefont {N.}~\bibnamefont {Chetty}},\ }\bibfield
  {title} {\enquote {\bibinfo {title} {Defect charge states in si doped
  hexagonal boron-nitride monolayer},}\ }\href
  {http://stacks.iop.org/0953-8984/28/i=5/a=055501} {\bibfield  {journal}
  {\bibinfo  {journal} {Journal of Physics: Condensed Matter}\ }\textbf
  {\bibinfo {volume} {28}},\ \bibinfo {pages} {055501} (\bibinfo {year}
  {2016})}\BibitemShut {NoStop}%
\bibitem [{\citenamefont {Jin}\ \emph {et~al.}(2009)\citenamefont {Jin},
  \citenamefont {Lin}, \citenamefont {Suenaga},\ and\ \citenamefont
  {Iijima}}]{Jin2009}%
  \BibitemOpen
  \bibfield  {author} {\bibinfo {author} {\bibfnamefont {C.}~\bibnamefont
  {Jin}}, \bibinfo {author} {\bibfnamefont {F.}~\bibnamefont {Lin}}, \bibinfo
  {author} {\bibfnamefont {K.}~\bibnamefont {Suenaga}}, \ and\ \bibinfo
  {author} {\bibfnamefont {S.}~\bibnamefont {Iijima}},\ }\bibfield  {title}
  {\enquote {\bibinfo {title} {Fabrication of a freestanding boron nitride
  single layer and its defect assignments},}\ }\href {\doibase
  10.1103/PhysRevLett.102.195505} {\bibfield  {journal} {\bibinfo  {journal}
  {Phys. Rev. Lett.}\ }\textbf {\bibinfo {volume} {102}},\ \bibinfo {pages}
  {195505} (\bibinfo {year} {2009})}\BibitemShut {NoStop}%
\bibitem [{\citenamefont {Kotakoski}\ \emph {et~al.}(2010)\citenamefont
  {Kotakoski}, \citenamefont {Jin}, \citenamefont {Lehtinen}, \citenamefont
  {Suenaga},\ and\ \citenamefont {Krasheninnikov}}]{Kotakoski2010}%
  \BibitemOpen
  \bibfield  {author} {\bibinfo {author} {\bibfnamefont {J.}~\bibnamefont
  {Kotakoski}}, \bibinfo {author} {\bibfnamefont {C.~H.}\ \bibnamefont {Jin}},
  \bibinfo {author} {\bibfnamefont {O.}~\bibnamefont {Lehtinen}}, \bibinfo
  {author} {\bibfnamefont {K.}~\bibnamefont {Suenaga}}, \ and\ \bibinfo
  {author} {\bibfnamefont {A.~V.}\ \bibnamefont {Krasheninnikov}},\ }\bibfield
  {title} {\enquote {\bibinfo {title} {Electron knock-on damage in hexagonal
  boron nitride monolayers},}\ }\href {\doibase 10.1103/PhysRevB.82.113404}
  {\bibfield  {journal} {\bibinfo  {journal} {Phys. Rev. B}\ }\textbf {\bibinfo
  {volume} {82}},\ \bibinfo {pages} {113404} (\bibinfo {year}
  {2010})}\BibitemShut {NoStop}%
\bibitem [{\citenamefont {Krivanek}\ \emph {et~al.}(2008)\citenamefont
  {Krivanek}, \citenamefont {Corbin}, \citenamefont {Dellby}, \citenamefont
  {Elston}, \citenamefont {Keyse}, \citenamefont {Murfitt}, \citenamefont
  {Own}, \citenamefont {Szilagyi},\ and\ \citenamefont
  {Woodruff}}]{krivanek_electron_2008}%
  \BibitemOpen
  \bibfield  {author} {\bibinfo {author} {\bibfnamefont {O.~L.}\ \bibnamefont
  {Krivanek}}, \bibinfo {author} {\bibfnamefont {G.~J.}\ \bibnamefont
  {Corbin}}, \bibinfo {author} {\bibfnamefont {N.}~\bibnamefont {Dellby}},
  \bibinfo {author} {\bibfnamefont {B.~F.}\ \bibnamefont {Elston}}, \bibinfo
  {author} {\bibfnamefont {R.~J.}\ \bibnamefont {Keyse}}, \bibinfo {author}
  {\bibfnamefont {M.~F.}\ \bibnamefont {Murfitt}}, \bibinfo {author}
  {\bibfnamefont {C.~S.}\ \bibnamefont {Own}}, \bibinfo {author} {\bibfnamefont
  {Z.~S.}\ \bibnamefont {Szilagyi}}, \ and\ \bibinfo {author} {\bibfnamefont
  {J.~W.}\ \bibnamefont {Woodruff}},\ }\bibfield  {title} {\enquote {\bibinfo
  {title} {An electron microscope for the aberration-corrected era},}\ }\href
  {\doibase 10.1016/j.ultramic.2007.07.010} {\bibfield  {journal} {\bibinfo
  {journal} {Ultramicroscopy}\ }\textbf {\bibinfo {volume} {108}},\ \bibinfo
  {pages} {179--195} (\bibinfo {year} {2008})}\BibitemShut {NoStop}%
\bibitem [{\citenamefont {Leuthner}\ \emph {et~al.}(2018)\citenamefont
  {Leuthner}, \citenamefont {Hummel}, \citenamefont {Mangler}, \citenamefont
  {Pennycook}, \citenamefont {Susi}, \citenamefont {Meyer},\ and\ \citenamefont
  {Kotakoski}}]{leuthner_scanning_2018}%
  \BibitemOpen
  \bibfield  {author} {\bibinfo {author} {\bibfnamefont {G.~T.}\ \bibnamefont
  {Leuthner}}, \bibinfo {author} {\bibfnamefont {S.}~\bibnamefont {Hummel}},
  \bibinfo {author} {\bibfnamefont {C.}~\bibnamefont {Mangler}}, \bibinfo
  {author} {\bibfnamefont {T.~J.}\ \bibnamefont {Pennycook}}, \bibinfo {author}
  {\bibfnamefont {T.}~\bibnamefont {Susi}}, \bibinfo {author} {\bibfnamefont
  {J.~C.}\ \bibnamefont {Meyer}}, \ and\ \bibinfo {author} {\bibfnamefont
  {J.}~\bibnamefont {Kotakoski}},\ }\bibfield  {title} {\enquote {\bibinfo
  {title} {Scanning transmission electron microscopy under controlled
  low-pressure atmospheres},}\ }\href {http://arxiv.org/abs/1811.04266}
  {\bibfield  {journal} {\bibinfo  {journal} {arXiv:1811.04266 [cond-mat]}\ }
  (\bibinfo {year} {2018})},\ \bibinfo {note} {arXiv: 1811.04266}\BibitemShut
  {NoStop}%
\bibitem [{\citenamefont {Koch}(2002)}]{Koch2002}%
  \BibitemOpen
  \bibfield  {author} {\bibinfo {author} {\bibfnamefont {C.}~\bibnamefont
  {Koch}},\ }\emph {\bibinfo {title} {Determination of core structure
  periodicity and point defect density along dislocations PhD Thesis Arizona
  State University}},\ \href@noop {} {Ph.D. thesis},\ \bibinfo  {school}
  {Arizona State University} (\bibinfo {year} {2002})\BibitemShut {NoStop}%
\bibitem [{\citenamefont {Ramasse}\ \emph
  {et~al.}(2013{\natexlab{b}})\citenamefont {Ramasse}, \citenamefont
  {Seabourne}, \citenamefont {Kepaptsoglou}, \citenamefont {Zan}, \citenamefont
  {Bangert},\ and\ \citenamefont {Scott}}]{ramasse_probing_2013}%
  \BibitemOpen
  \bibfield  {author} {\bibinfo {author} {\bibfnamefont {Q.~M.}\ \bibnamefont
  {Ramasse}}, \bibinfo {author} {\bibfnamefont {C.~R.}\ \bibnamefont
  {Seabourne}}, \bibinfo {author} {\bibfnamefont {D.-M.}\ \bibnamefont
  {Kepaptsoglou}}, \bibinfo {author} {\bibfnamefont {R.}~\bibnamefont {Zan}},
  \bibinfo {author} {\bibfnamefont {U.}~\bibnamefont {Bangert}}, \ and\
  \bibinfo {author} {\bibfnamefont {A.~J.}\ \bibnamefont {Scott}},\ }\bibfield
  {title} {\enquote {\bibinfo {title} {Probing the {Bonding} and {Electronic}
  {Structure} of {Single} {Atom} {Dopants} in {Graphene} with {Electron}
  {Energy} {Loss} {Spectroscopy}},}\ }\href {\doibase 10.1021/nl304187e}
  {\bibfield  {journal} {\bibinfo  {journal} {Nano Lett.}\ }\textbf {\bibinfo
  {volume} {13}},\ \bibinfo {pages} {4989--4995} (\bibinfo {year}
  {2013}{\natexlab{b}})}\BibitemShut {NoStop}%
\bibitem [{\citenamefont {Inani}\ \emph {et~al.}(2019)\citenamefont {Inani},
  \citenamefont {Mustonen}, \citenamefont {Markevich}, \citenamefont {Ding},
  \citenamefont {Tripathi}, \citenamefont {Hussain}, \citenamefont {Mangler},
  \citenamefont {Kauppinen}, \citenamefont {Susi},\ and\ \citenamefont
  {Kotakoski}}]{Inani2019}%
  \BibitemOpen
  \bibfield  {author} {\bibinfo {author} {\bibfnamefont {H.}~\bibnamefont
  {Inani}}, \bibinfo {author} {\bibfnamefont {K.}~\bibnamefont {Mustonen}},
  \bibinfo {author} {\bibfnamefont {A.}~\bibnamefont {Markevich}}, \bibinfo
  {author} {\bibfnamefont {E.-X.}\ \bibnamefont {Ding}}, \bibinfo {author}
  {\bibfnamefont {M.}~\bibnamefont {Tripathi}}, \bibinfo {author}
  {\bibfnamefont {A.}~\bibnamefont {Hussain}}, \bibinfo {author} {\bibfnamefont
  {C.}~\bibnamefont {Mangler}}, \bibinfo {author} {\bibfnamefont {E.~I.}\
  \bibnamefont {Kauppinen}}, \bibinfo {author} {\bibfnamefont {T.}~\bibnamefont
  {Susi}}, \ and\ \bibinfo {author} {\bibfnamefont {J.}~\bibnamefont
  {Kotakoski}},\ }\bibfield  {title} {\enquote {\bibinfo {title} {Silicon
  substitution in nanotubes and graphene via intermittent vacancies},}\ }\href
  {\doibase 10.1021/acs.jpcc.9b01894} {\bibfield  {journal} {\bibinfo
  {journal} {The Journal of Physical Chemistry C}\ }\textbf {\bibinfo {volume}
  {123}},\ \bibinfo {pages} {13136--13140} (\bibinfo {year} {2019})},\ \Eprint
  {http://arxiv.org/abs/https://doi.org/10.1021/acs.jpcc.9b01894}
  {https://doi.org/10.1021/acs.jpcc.9b01894} \BibitemShut {NoStop}%
\bibitem [{\citenamefont {Kresse}\ and\ \citenamefont
  {Hafner}(1993)}]{Kresse1993}%
  \BibitemOpen
  \bibfield  {author} {\bibinfo {author} {\bibfnamefont {G.}~\bibnamefont
  {Kresse}}\ and\ \bibinfo {author} {\bibfnamefont {J.}~\bibnamefont
  {Hafner}},\ }\bibfield  {title} {\enquote {\bibinfo {title} {Ab initio
  molecular dynamics for liquid metals},}\ }\href {\doibase
  10.1103/PhysRevB.47.558} {\bibfield  {journal} {\bibinfo  {journal} {Phys.
  Rev. B}\ }\textbf {\bibinfo {volume} {47}},\ \bibinfo {pages} {558--561}
  (\bibinfo {year} {1993})}\BibitemShut {NoStop}%
\bibitem [{\citenamefont {Perdew}, \citenamefont {Burke},\ and\ \citenamefont
  {Ernzerhof}(1996)}]{Perdew1996}%
  \BibitemOpen
  \bibfield  {author} {\bibinfo {author} {\bibfnamefont {J.~P.}\ \bibnamefont
  {Perdew}}, \bibinfo {author} {\bibfnamefont {K.}~\bibnamefont {Burke}}, \
  and\ \bibinfo {author} {\bibfnamefont {M.}~\bibnamefont {Ernzerhof}},\
  }\bibfield  {title} {\enquote {\bibinfo {title} {Generalized gradient
  approximation made simple},}\ }\href {\doibase 10.1103/PhysRevLett.77.3865}
  {\bibfield  {journal} {\bibinfo  {journal} {Phys. Rev. Lett.}\ }\textbf
  {\bibinfo {volume} {77}},\ \bibinfo {pages} {3865--3868} (\bibinfo {year}
  {1996})}\BibitemShut {NoStop}%
\bibitem [{\citenamefont {Kresse}\ and\ \citenamefont
  {Joubert}(1999)}]{Kresse1999}%
  \BibitemOpen
  \bibfield  {author} {\bibinfo {author} {\bibfnamefont {G.}~\bibnamefont
  {Kresse}}\ and\ \bibinfo {author} {\bibfnamefont {D.}~\bibnamefont
  {Joubert}},\ }\bibfield  {title} {\enquote {\bibinfo {title} {From ultrasoft
  pseudopotentials to the projector augmented-wave method},}\ }\href {\doibase
  10.1103/PhysRevB.59.1758} {\bibfield  {journal} {\bibinfo  {journal} {Phys.
  Rev. B}\ }\textbf {\bibinfo {volume} {59}},\ \bibinfo {pages} {1758--1775}
  (\bibinfo {year} {1999})}\BibitemShut {NoStop}%
\bibitem [{\citenamefont {Komsa}\ \emph {et~al.}(2014)\citenamefont {Komsa},
  \citenamefont {Berseneva}, \citenamefont {Krasheninnikov},\ and\
  \citenamefont {Nieminen}}]{komsa2014}%
  \BibitemOpen
  \bibfield  {author} {\bibinfo {author} {\bibfnamefont {H.-P.}\ \bibnamefont
  {Komsa}}, \bibinfo {author} {\bibfnamefont {N.}~\bibnamefont {Berseneva}},
  \bibinfo {author} {\bibfnamefont {A.~V.}\ \bibnamefont {Krasheninnikov}}, \
  and\ \bibinfo {author} {\bibfnamefont {R.~M.}\ \bibnamefont {Nieminen}},\
  }\bibfield  {title} {\enquote {\bibinfo {title} {Charged point defects in the
  flatland: Accurate formation energy calculations in two-dimensional
  materials},}\ }\href {\doibase 10.1103/PhysRevX.4.031044} {\bibfield
  {journal} {\bibinfo  {journal} {Phys. Rev. X}\ }\textbf {\bibinfo {volume}
  {4}},\ \bibinfo {pages} {031044} (\bibinfo {year} {2014})}\BibitemShut
  {NoStop}%
\bibitem [{\citenamefont {Komsa}\ \emph {et~al.}(2018)\citenamefont {Komsa},
  \citenamefont {Berseneva}, \citenamefont {Krasheninnikov},\ and\
  \citenamefont {Nieminen}}]{Komsa2018}%
  \BibitemOpen
  \bibfield  {author} {\bibinfo {author} {\bibfnamefont {H.-P.}\ \bibnamefont
  {Komsa}}, \bibinfo {author} {\bibfnamefont {N.}~\bibnamefont {Berseneva}},
  \bibinfo {author} {\bibfnamefont {A.~V.}\ \bibnamefont {Krasheninnikov}}, \
  and\ \bibinfo {author} {\bibfnamefont {R.~M.}\ \bibnamefont {Nieminen}},\
  }\bibfield  {title} {\enquote {\bibinfo {title} {Erratum: Charged point
  defects in the flatland: Accurate formation energy calculations in
  two-dimensional materials [phys. rev. x 4, 031044 (2014)]},}\ }\href
  {\doibase 10.1103/PhysRevX.8.039902} {\bibfield  {journal} {\bibinfo
  {journal} {Phys. Rev. X}\ }\textbf {\bibinfo {volume} {8}},\ \bibinfo {pages}
  {039902} (\bibinfo {year} {2018})}\BibitemShut {NoStop}%
\bibitem [{\citenamefont {Heyd}, \citenamefont {Scuseria},\ and\ \citenamefont
  {Ernzerhof}(2003)}]{Heyd2003}%
  \BibitemOpen
  \bibfield  {author} {\bibinfo {author} {\bibfnamefont {J.}~\bibnamefont
  {Heyd}}, \bibinfo {author} {\bibfnamefont {G.~E.}\ \bibnamefont {Scuseria}},
  \ and\ \bibinfo {author} {\bibfnamefont {M.}~\bibnamefont {Ernzerhof}},\
  }\bibfield  {title} {\enquote {\bibinfo {title} {Hybrid functionals based on
  a screened coulomb potential},}\ }\href {\doibase 10.1063/1.1564060}
  {\bibfield  {journal} {\bibinfo  {journal} {The Journal of Chemical Physics}\
  }\textbf {\bibinfo {volume} {118}},\ \bibinfo {pages} {8207--8215} (\bibinfo
  {year} {2003})}\BibitemShut {NoStop}%
\bibitem [{\citenamefont {Berseneva}, \citenamefont {Krasheninnikov},\ and\
  \citenamefont {Nieminen}(2011)}]{Berseneva2011}%
  \BibitemOpen
  \bibfield  {author} {\bibinfo {author} {\bibfnamefont {N.}~\bibnamefont
  {Berseneva}}, \bibinfo {author} {\bibfnamefont {A.~V.}\ \bibnamefont
  {Krasheninnikov}}, \ and\ \bibinfo {author} {\bibfnamefont {R.~M.}\
  \bibnamefont {Nieminen}},\ }\bibfield  {title} {\enquote {\bibinfo {title}
  {Mechanisms of postsynthesis doping of boron nitride nanostructures with
  carbon from first-principles simulations},}\ }\href {\doibase
  10.1103/PhysRevLett.107.035501} {\bibfield  {journal} {\bibinfo  {journal}
  {Phys. Rev. Lett.}\ }\textbf {\bibinfo {volume} {107}},\ \bibinfo {pages}
  {035501} (\bibinfo {year} {2011})}\BibitemShut {NoStop}%
\bibitem [{\citenamefont {Krukau}\ \emph {et~al.}(2006)\citenamefont {Krukau},
  \citenamefont {Vydrov}, \citenamefont {Izmaylov},\ and\ \citenamefont
  {Scuseria}}]{Aliaksandr2006}%
  \BibitemOpen
  \bibfield  {author} {\bibinfo {author} {\bibfnamefont {A.~V.}\ \bibnamefont
  {Krukau}}, \bibinfo {author} {\bibfnamefont {O.~A.}\ \bibnamefont {Vydrov}},
  \bibinfo {author} {\bibfnamefont {A.~F.}\ \bibnamefont {Izmaylov}}, \ and\
  \bibinfo {author} {\bibfnamefont {G.~E.}\ \bibnamefont {Scuseria}},\
  }\bibfield  {title} {\enquote {\bibinfo {title} {Influence of the exchange
  screening parameter on the performance of screened hybrid functionals},}\
  }\href {\doibase 10.1063/1.2404663} {\bibfield  {journal} {\bibinfo
  {journal} {The Journal of Chemical Physics}\ }\textbf {\bibinfo {volume}
  {125}},\ \bibinfo {pages} {224106} (\bibinfo {year} {2006})}\BibitemShut
  {NoStop}%
\bibitem [{\citenamefont {Huang}\ and\ \citenamefont {Wei}(2011)}]{Huang2011}%
  \BibitemOpen
  \bibfield  {author} {\bibinfo {author} {\bibfnamefont {B.}~\bibnamefont
  {Huang}}\ and\ \bibinfo {author} {\bibfnamefont {S.-H.}\ \bibnamefont
  {Wei}},\ }\bibfield  {title} {\enquote {\bibinfo {title} {Comment on
  ``mechanisms of postsynthesis doping of boron nitride nanostructures with
  carbon from first-principles simulations''},}\ }\href {\doibase
  10.1103/PhysRevLett.107.239601} {\bibfield  {journal} {\bibinfo  {journal}
  {Phys. Rev. Lett.}\ }\textbf {\bibinfo {volume} {107}},\ \bibinfo {pages}
  {239601} (\bibinfo {year} {2011})}\BibitemShut {NoStop}%
\bibitem [{\citenamefont {Huang}\ and\ \citenamefont {Lee}(2012)}]{Huang2012}%
  \BibitemOpen
  \bibfield  {author} {\bibinfo {author} {\bibfnamefont {B.}~\bibnamefont
  {Huang}}\ and\ \bibinfo {author} {\bibfnamefont {H.}~\bibnamefont {Lee}},\
  }\bibfield  {title} {\enquote {\bibinfo {title} {Defect and impurity
  properties of hexagonal boron nitride: A first-principles calculation},}\
  }\href {\doibase 10.1103/PhysRevB.86.245406} {\bibfield  {journal} {\bibinfo
  {journal} {Phys. Rev. B}\ }\textbf {\bibinfo {volume} {86}},\ \bibinfo
  {pages} {245406} (\bibinfo {year} {2012})}\BibitemShut {NoStop}%
\bibitem [{\citenamefont {Van~de Walle}\ \emph {et~al.}(1989)\citenamefont
  {Van~de Walle}, \citenamefont {Denteneer}, \citenamefont {Bar-Yam},\ and\
  \citenamefont {Pantelides}}]{Walle1989}%
  \BibitemOpen
  \bibfield  {author} {\bibinfo {author} {\bibfnamefont {C.~G.}\ \bibnamefont
  {Van~de Walle}}, \bibinfo {author} {\bibfnamefont {P.~J.~H.}\ \bibnamefont
  {Denteneer}}, \bibinfo {author} {\bibfnamefont {Y.}~\bibnamefont {Bar-Yam}},
  \ and\ \bibinfo {author} {\bibfnamefont {S.~T.}\ \bibnamefont {Pantelides}},\
  }\bibfield  {title} {\enquote {\bibinfo {title} {Theory of hydrogen diffusion
  and reactions in crystalline silicon},}\ }\href {\doibase
  10.1103/PhysRevB.39.10791} {\bibfield  {journal} {\bibinfo  {journal} {Phys.
  Rev. B}\ }\textbf {\bibinfo {volume} {39}},\ \bibinfo {pages} {10791--10808}
  (\bibinfo {year} {1989})}\BibitemShut {NoStop}%
\bibitem [{\citenamefont {van Setten}\ \emph {et~al.}(2007)\citenamefont {van
  Setten}, \citenamefont {Uijttewaal}, \citenamefont {de~Wijs},\ and\
  \citenamefont {de~Groot}}]{Setten2007}%
  \BibitemOpen
  \bibfield  {author} {\bibinfo {author} {\bibfnamefont {M.~J.}\ \bibnamefont
  {van Setten}}, \bibinfo {author} {\bibfnamefont {M.~A.}\ \bibnamefont
  {Uijttewaal}}, \bibinfo {author} {\bibfnamefont {G.~A.}\ \bibnamefont
  {de~Wijs}}, \ and\ \bibinfo {author} {\bibfnamefont {R.~A.}\ \bibnamefont
  {de~Groot}},\ }\bibfield  {title} {\enquote {\bibinfo {title} {Thermodynamic
  stability of boron:the role of defects and zero point motion},}\ }\href
  {\doibase 10.1021/ja0631246} {\bibfield  {journal} {\bibinfo  {journal}
  {Journal of the American Chemical Society}\ }\textbf {\bibinfo {volume}
  {129}},\ \bibinfo {pages} {2458--2465} (\bibinfo {year} {2007})}\BibitemShut
  {NoStop}%
\bibitem [{\citenamefont {Laskowski}, \citenamefont {Blaha},\ and\
  \citenamefont {Schwarz}(2008)}]{Laskowski2008}%
  \BibitemOpen
  \bibfield  {author} {\bibinfo {author} {\bibfnamefont {R.}~\bibnamefont
  {Laskowski}}, \bibinfo {author} {\bibfnamefont {P.}~\bibnamefont {Blaha}}, \
  and\ \bibinfo {author} {\bibfnamefont {K.}~\bibnamefont {Schwarz}},\
  }\bibfield  {title} {\enquote {\bibinfo {title} {Bonding of hexagonal bn to
  transition metal surfaces: An ab initio density-functional theory study},}\
  }\href {\doibase 10.1103/PhysRevB.78.045409} {\bibfield  {journal} {\bibinfo
  {journal} {Phys. Rev. B}\ }\textbf {\bibinfo {volume} {78}},\ \bibinfo
  {pages} {045409} (\bibinfo {year} {2008})}\BibitemShut {NoStop}%
\bibitem [{\citenamefont {Zhang}\ \emph {et~al.}(2018)\citenamefont {Zhang},
  \citenamefont {Yu}, \citenamefont {Ebert}, \citenamefont {Zhang},
  \citenamefont {Pan}, \citenamefont {Chou}, \citenamefont {Shih},
  \citenamefont {Zeng},\ and\ \citenamefont {Yuan}}]{Zhang2018}%
  \BibitemOpen
  \bibfield  {author} {\bibinfo {author} {\bibfnamefont {Q.}~\bibnamefont
  {Zhang}}, \bibinfo {author} {\bibfnamefont {J.}~\bibnamefont {Yu}}, \bibinfo
  {author} {\bibfnamefont {P.}~\bibnamefont {Ebert}}, \bibinfo {author}
  {\bibfnamefont {C.}~\bibnamefont {Zhang}}, \bibinfo {author} {\bibfnamefont
  {C.-R.}\ \bibnamefont {Pan}}, \bibinfo {author} {\bibfnamefont {M.-Y.}\
  \bibnamefont {Chou}}, \bibinfo {author} {\bibfnamefont {C.-K.}\ \bibnamefont
  {Shih}}, \bibinfo {author} {\bibfnamefont {C.}~\bibnamefont {Zeng}}, \ and\
  \bibinfo {author} {\bibfnamefont {S.}~\bibnamefont {Yuan}},\ }\bibfield
  {title} {\enquote {\bibinfo {title} {Tuning band gap and work function
  modulations in monolayer hbn/cu(111) heterostructures with moiré
  patterns},}\ }\href {\doibase 10.1021/acsnano.8b04444} {\bibfield  {journal}
  {\bibinfo  {journal} {ACS Nano}\ }\textbf {\bibinfo {volume} {12}},\ \bibinfo
  {pages} {9355--9362} (\bibinfo {year} {2018})}\BibitemShut {NoStop}%
\bibitem [{\citenamefont {Wei}\ \emph {et~al.}(2013)\citenamefont {Wei},
  \citenamefont {Tang}, \citenamefont {Chen}, \citenamefont {Bando},\ and\
  \citenamefont {Golberg}}]{Wei2013}%
  \BibitemOpen
  \bibfield  {author} {\bibinfo {author} {\bibfnamefont {X.}~\bibnamefont
  {Wei}}, \bibinfo {author} {\bibfnamefont {D.-M.}\ \bibnamefont {Tang}},
  \bibinfo {author} {\bibfnamefont {Q.}~\bibnamefont {Chen}}, \bibinfo {author}
  {\bibfnamefont {Y.}~\bibnamefont {Bando}}, \ and\ \bibinfo {author}
  {\bibfnamefont {D.}~\bibnamefont {Golberg}},\ }\bibfield  {title} {\enquote
  {\bibinfo {title} {Local coulomb explosion of boron nitride nanotubes under
  electron beam irradiation},}\ }\href {\doibase 10.1021/nn400423y} {\bibfield
  {journal} {\bibinfo  {journal} {ACS Nano}\ }\textbf {\bibinfo {volume} {7}},\
  \bibinfo {pages} {3491--3497} (\bibinfo {year} {2013})}\BibitemShut {NoStop}%
\bibitem [{\citenamefont {Hofer}\ \emph {et~al.}(2019)\citenamefont {Hofer},
  \citenamefont {Skakalova}, \citenamefont {Monazam}, \citenamefont {Mangler},
  \citenamefont {Kotakoski}, \citenamefont {Susi},\ and\ \citenamefont
  {Meyer}}]{hofer_direct_2019}%
  \BibitemOpen
  \bibfield  {author} {\bibinfo {author} {\bibfnamefont {C.}~\bibnamefont
  {Hofer}}, \bibinfo {author} {\bibfnamefont {V.}~\bibnamefont {Skakalova}},
  \bibinfo {author} {\bibfnamefont {M.~R.~A.}\ \bibnamefont {Monazam}},
  \bibinfo {author} {\bibfnamefont {C.}~\bibnamefont {Mangler}}, \bibinfo
  {author} {\bibfnamefont {J.}~\bibnamefont {Kotakoski}}, \bibinfo {author}
  {\bibfnamefont {T.}~\bibnamefont {Susi}}, \ and\ \bibinfo {author}
  {\bibfnamefont {J.~C.}\ \bibnamefont {Meyer}},\ }\bibfield  {title} {\enquote
  {\bibinfo {title} {Direct visualization of the 3d structure of silicon
  impurities in graphene},}\ }\href {\doibase 10.1063/1.5063449} {\bibfield
  {journal} {\bibinfo  {journal} {Appl. Phys. Lett.}\ }\textbf {\bibinfo
  {volume} {114}},\ \bibinfo {pages} {053102} (\bibinfo {year}
  {2019})}\BibitemShut {NoStop}%
\bibitem [{\citenamefont {Henkelman}\ and\ \citenamefont
  {J\'onsson}(2000)}]{Henkelman2000}%
  \BibitemOpen
  \bibfield  {author} {\bibinfo {author} {\bibfnamefont {G.}~\bibnamefont
  {Henkelman}}\ and\ \bibinfo {author} {\bibfnamefont {H.}~\bibnamefont
  {J\'onsson}},\ }\bibfield  {title} {\enquote {\bibinfo {title} {Improved
  tangent estimate in the nudged elastic band method for finding minimum energy
  paths and saddle points},}\ }\href {\doibase 10.1063/1.1323224} {\bibfield
  {journal} {\bibinfo  {journal} {The Journal of Chemical Physics}\ }\textbf
  {\bibinfo {volume} {113}},\ \bibinfo {pages} {9978--9985} (\bibinfo {year}
  {2000})}\BibitemShut {NoStop}%
\bibitem [{\citenamefont {Susi}\ \emph {et~al.}(2014)\citenamefont {Susi},
  \citenamefont {Kotakoski}, \citenamefont {Kepaptsoglou}, \citenamefont
  {Mangler}, \citenamefont {Lovejoy}, \citenamefont {Krivanek}, \citenamefont
  {Zan}, \citenamefont {Bangert}, \citenamefont {Ayala}, \citenamefont
  {Meyer},\ and\ \citenamefont {Ramasse}}]{Susi2014}%
  \BibitemOpen
  \bibfield  {author} {\bibinfo {author} {\bibfnamefont {T.}~\bibnamefont
  {Susi}}, \bibinfo {author} {\bibfnamefont {J.}~\bibnamefont {Kotakoski}},
  \bibinfo {author} {\bibfnamefont {D.}~\bibnamefont {Kepaptsoglou}}, \bibinfo
  {author} {\bibfnamefont {C.}~\bibnamefont {Mangler}}, \bibinfo {author}
  {\bibfnamefont {T.~C.}\ \bibnamefont {Lovejoy}}, \bibinfo {author}
  {\bibfnamefont {O.~L.}\ \bibnamefont {Krivanek}}, \bibinfo {author}
  {\bibfnamefont {R.}~\bibnamefont {Zan}}, \bibinfo {author} {\bibfnamefont
  {U.}~\bibnamefont {Bangert}}, \bibinfo {author} {\bibfnamefont
  {P.}~\bibnamefont {Ayala}}, \bibinfo {author} {\bibfnamefont {J.~C.}\
  \bibnamefont {Meyer}}, \ and\ \bibinfo {author} {\bibfnamefont
  {Q.}~\bibnamefont {Ramasse}},\ }\bibfield  {title} {\enquote {\bibinfo
  {title} {Silicon--carbon bond inversions driven by 60-kev electrons in
  graphene},}\ }\href {\doibase 10.1103/PhysRevLett.113.115501} {\bibfield
  {journal} {\bibinfo  {journal} {Phys. Rev. Lett.}\ }\textbf {\bibinfo
  {volume} {113}},\ \bibinfo {pages} {115501} (\bibinfo {year}
  {2014})}\BibitemShut {NoStop}%
\bibitem [{\citenamefont {Susi}\ \emph {et~al.}(2017)\citenamefont {Susi},
  \citenamefont {Hardcastle}, \citenamefont {Hofsäss}, \citenamefont
  {Mittelberger}, \citenamefont {Pennycook}, \citenamefont {Mangler},
  \citenamefont {Drummond-Brydson}, \citenamefont {Scott}, \citenamefont
  {Meyer},\ and\ \citenamefont {Kotakoski}}]{Susi2017}%
  \BibitemOpen
  \bibfield  {author} {\bibinfo {author} {\bibfnamefont {T.}~\bibnamefont
  {Susi}}, \bibinfo {author} {\bibfnamefont {T.~P.}\ \bibnamefont
  {Hardcastle}}, \bibinfo {author} {\bibfnamefont {H.}~\bibnamefont
  {Hofsäss}}, \bibinfo {author} {\bibfnamefont {A.}~\bibnamefont
  {Mittelberger}}, \bibinfo {author} {\bibfnamefont {T.~J.}\ \bibnamefont
  {Pennycook}}, \bibinfo {author} {\bibfnamefont {C.}~\bibnamefont {Mangler}},
  \bibinfo {author} {\bibfnamefont {R.}~\bibnamefont {Drummond-Brydson}},
  \bibinfo {author} {\bibfnamefont {A.~J.}\ \bibnamefont {Scott}}, \bibinfo
  {author} {\bibfnamefont {J.~C.}\ \bibnamefont {Meyer}}, \ and\ \bibinfo
  {author} {\bibfnamefont {J.}~\bibnamefont {Kotakoski}},\ }\bibfield  {title}
  {\enquote {\bibinfo {title} {Single-atom spectroscopy of phosphorus dopants
  implanted into graphene},}\ }\href
  {http://stacks.iop.org/2053-1583/4/i=2/a=021013} {\bibfield  {journal}
  {\bibinfo  {journal} {2D Materials}\ }\textbf {\bibinfo {volume} {4}},\
  \bibinfo {pages} {021013} (\bibinfo {year} {2017})}\BibitemShut {NoStop}%
\bibitem [{\citenamefont {Tripathi}\ \emph {et~al.}(2018)\citenamefont
  {Tripathi}, \citenamefont {Markevich}, \citenamefont {Böttger},
  \citenamefont {Facsko}, \citenamefont {Besley}, \citenamefont {Kotakoski},\
  and\ \citenamefont {Susi}}]{Tripathi2018}%
  \BibitemOpen
  \bibfield  {author} {\bibinfo {author} {\bibfnamefont {M.}~\bibnamefont
  {Tripathi}}, \bibinfo {author} {\bibfnamefont {A.}~\bibnamefont {Markevich}},
  \bibinfo {author} {\bibfnamefont {R.}~\bibnamefont {Böttger}}, \bibinfo
  {author} {\bibfnamefont {S.}~\bibnamefont {Facsko}}, \bibinfo {author}
  {\bibfnamefont {E.}~\bibnamefont {Besley}}, \bibinfo {author} {\bibfnamefont
  {J.}~\bibnamefont {Kotakoski}}, \ and\ \bibinfo {author} {\bibfnamefont
  {T.}~\bibnamefont {Susi}},\ }\bibfield  {title} {\enquote {\bibinfo {title}
  {Implanting germanium into graphene},}\ }\href {\doibase
  10.1021/acsnano.8b01191} {\bibfield  {journal} {\bibinfo  {journal} {ACS
  Nano}\ }\textbf {\bibinfo {volume} {12}},\ \bibinfo {pages} {4641--4647}
  (\bibinfo {year} {2018})}\BibitemShut {NoStop}%
\end{thebibliography}%

\hbox{}
\newpage

\section*{Supplementary Material}
\renewcommand{\thefigure}{S\arabic{figure}} 
\setcounter{figure}{0}

\title{Substitutional Si impurities in monolayer hexagonal boron nitride\\---\\Supplementary Material}

\author{Mohammad Reza Ahmadpour Monazam}
\email{mohammad.monazam@univie.ac.at}
\affiliation{University of Vienna, Faculty of Physics, Boltzmanngasse 5, A-1090, Vienna, Austria}
\author{Ursula Ludacka}
\affiliation{University of Vienna, Faculty of Physics, Boltzmanngasse 5, A-1090, Vienna, Austria}
\author{Hannu-Pekka Komsa}
\affiliation{Department of Applied Physics, Aalto University, P.O. Box 11100, 00076 Aalto, Finland}
\author{Jani Kotakoski}
\affiliation{University of Vienna, Faculty of Physics, Boltzmanngasse 5, A-1090, Vienna, Austria}

\date{\today}

\maketitle

\section*{Methods}

The samples were commercially available single-layer h-BN grown via chemical vapor deposition on copper by Graphene Laboratories, Inc. They were directly transferred onto golden transmission electron microscopy grids with perforated amorphous carbon membrane (QUANTIFOIL\textregistered) without the use of a polymer, which decreases the amount of contamination on the samples. The copper was etched in a bath of FeCl over night and the samples were cleaned with deionized water and isopropyl alcohol. Samples were baked in vacuum at 150$^\circ$C for at least eight hours before being inserted into the microscope.

We acquired the experimental data using a Nion UltraSTEM 100 microscope~\cite{krivanek_electron_2008} with a cold field-emission electron gun operated at 60~keV. The near-ultrahigh vacuum conditions at the objective area (pressure below $10^{-9}$~mbar) around the sample ensure a minimum influence of chemical reactions~\cite{leuthner_scanning_2018} on the sample during observation. The beam convergence semiangle was 30~mrad and the used medium angle annular dark field detector angular range 60-200~mrad. Typical beam current of the device is on the order of 30~pA.

We used density functional theory as implemented in the Vienna ab initio simulation package (VASP)~\cite{Kresse1993}. The electron exchange and correlation was treated by Perdew-Burke-Ernzerdorf (PBE) functional \cite{Perdew1996}. The total energy of the system was calculated via the pseudopotential-momentum-space formalism using projector-augmented-wave (PAW) method \cite{Kresse1999}. The Kohn-Sham wavefunctions are expanded over plane-wave basis sets with the kinetic energy cut off set to 525~eV. A supercell of $8\times8\times1$ was employed to study different defect states in the membrane with the assumption of minimizing the lateral interaction of the defect with its periodic images. The interlayer vacuum space of 43.46 \text{\AA} was selected according to "special vacuum" proposed in Refs. \cite{komsa2014, Komsa2018}. The results were compared to those calculated with a supercell of $6\times6\times1$ and vacuum size of 30.27 \text{\AA}. The locally optimized configurations and formation energies were in good agreement for the two different system sizes. The Brillouin-Zone integration was done over a $\Gamma$-centered $5\times5\times1$ k-point mesh. The damped molecular dynamics method was used to optimize the ionic degrees of freedom until the residual forces were below 0.01 eV/$\text{\AA}$. Although it is known that the band gaps calculated using PBE underestimate the true band gap of semiconductors, we restricted our calculation to the level of PBE due to the agreement between PBE formation energies and those calculated with the HSE formalism \cite{Heyd2003,Berseneva2011}. Therefore one only needs to re-scale the electron chemical potential using the difference in the band gap obtained from the two methods. Due to the computational cost, we carried out only one HSE calculation for bulk h-BN for estimating the band gap. The size of the band gap in this case is 5.72 eV as compared to 4.48 eV calculated with PBE. The HSE calculation is performed using HSE06 functional with 0.25 fraction of exact exchange \cite{Aliaksandr2006}. For STEM image simulations, we used the QSTEM package~\cite{Koch2002}, where all parameters were set to correspond to our experimental setup. The energy barrier estimation is based on the nudged elastic band (NEB) method implemented in VASP~\cite{Henkelman2000}. A set of calculations with five images between the initial and final configurations were performed. The standard dynamic calculation was performed using in DFT-based molecular dynamics with Nos\'e-thermostat ; an increasing initial vertical velocity toward the h-BN plane is applied to the silicon atom until it passed through the membrane. The time step is set to 0.5 fs.

\begin{figure}[h!]
	\includegraphics[width=.37\textwidth]{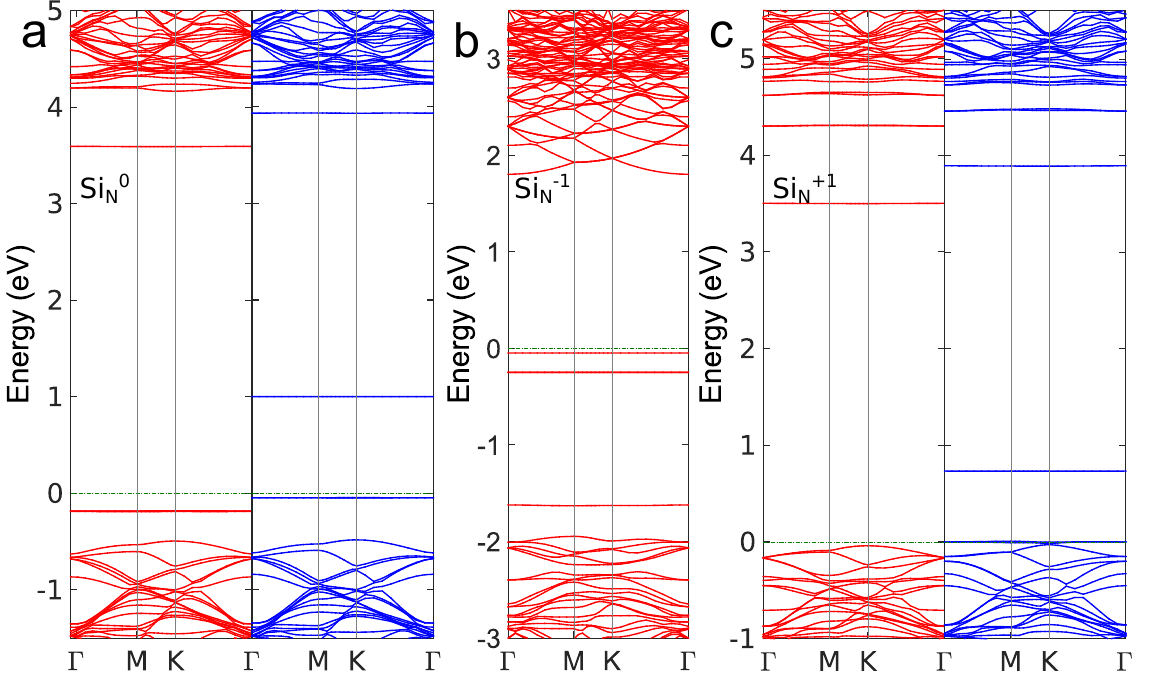}
	\caption{\label{Bands-SiN} Electronic band structure of Si$_N^0$, Si$_N^{-1}$ and Si$_N^{+1}$. The Fermi Level is set to zero. The first defect level in panel (a) around -0.18 eV in the spin up channel actually consists of three levels very close in energy, and the level close to Fermi level in the spin down channel is actually two levels. This enables charge states from -3 to +5. By adding an electron, to the lowest empty band in the spin down channel becomes spin-unpolarized (panel b). The defect level at 3.93 eV in the spin up channel and level 3.59 eV in the spin up channel are close to CBM and expected to become higher by adding an electron. In this case, the NFE bands push further down so that no further defect levels remain. Likewise, in +1 charge state, only one defect level very close to Fermi level remains.}
\end{figure}

\begin{figure}[h!]
	\includegraphics[width=.37\textwidth]{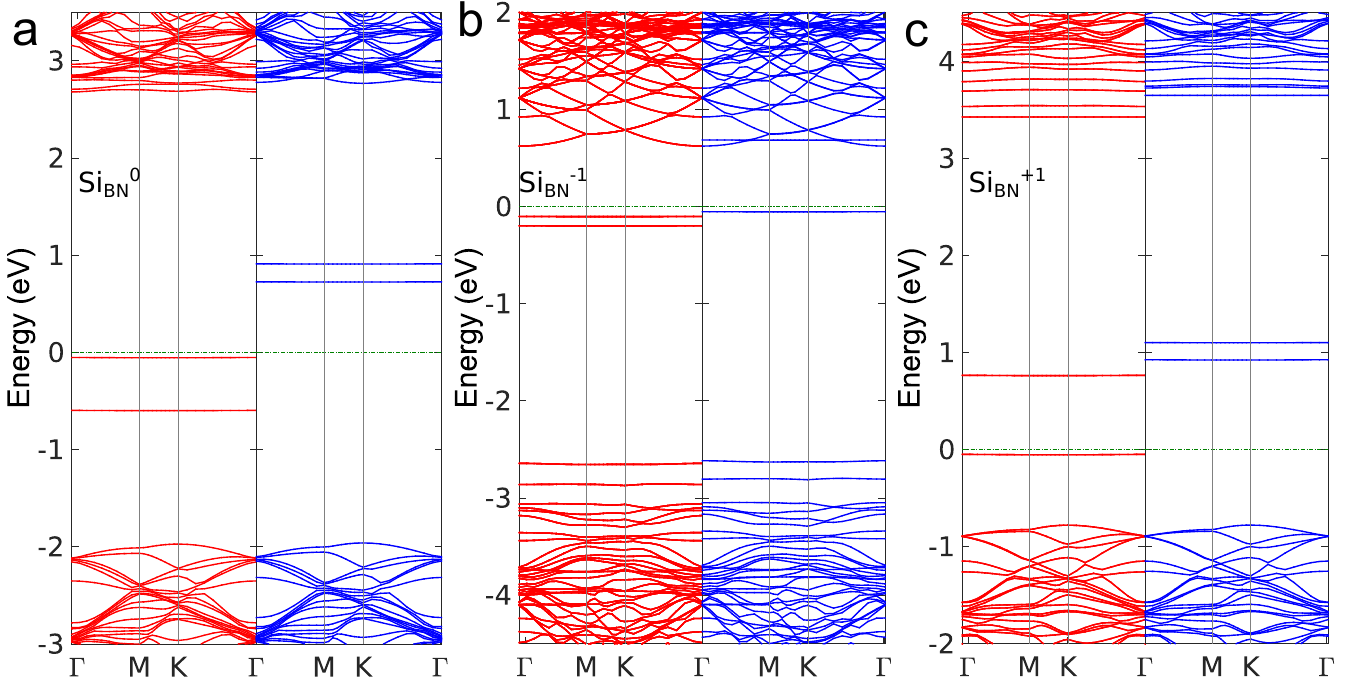}
	\caption{\label{SiDV-BS} lectronic band structures of Si$_{BN}^0$, Si$_{BN}^{-1}$ and Si$_{BN}^{+1}$. The band structure for neutral charge case (panel a) reveals four defect levels in the spin up and the spin down channels. Both occupied (empty) levels are in the spin up (spin down) channel. Therefore, the possible charge states extend from -2 to +2. However, from band structure of the -1 charge state (panel b), it is clear that adding a single electron to the lowest defect level in the spin down channel pushes the second defect level higher than CBM. So, only -1 state should be possible.}
\end{figure}

\begin{figure}[h!]
    \centering
    \includegraphics[width=.37\textwidth]{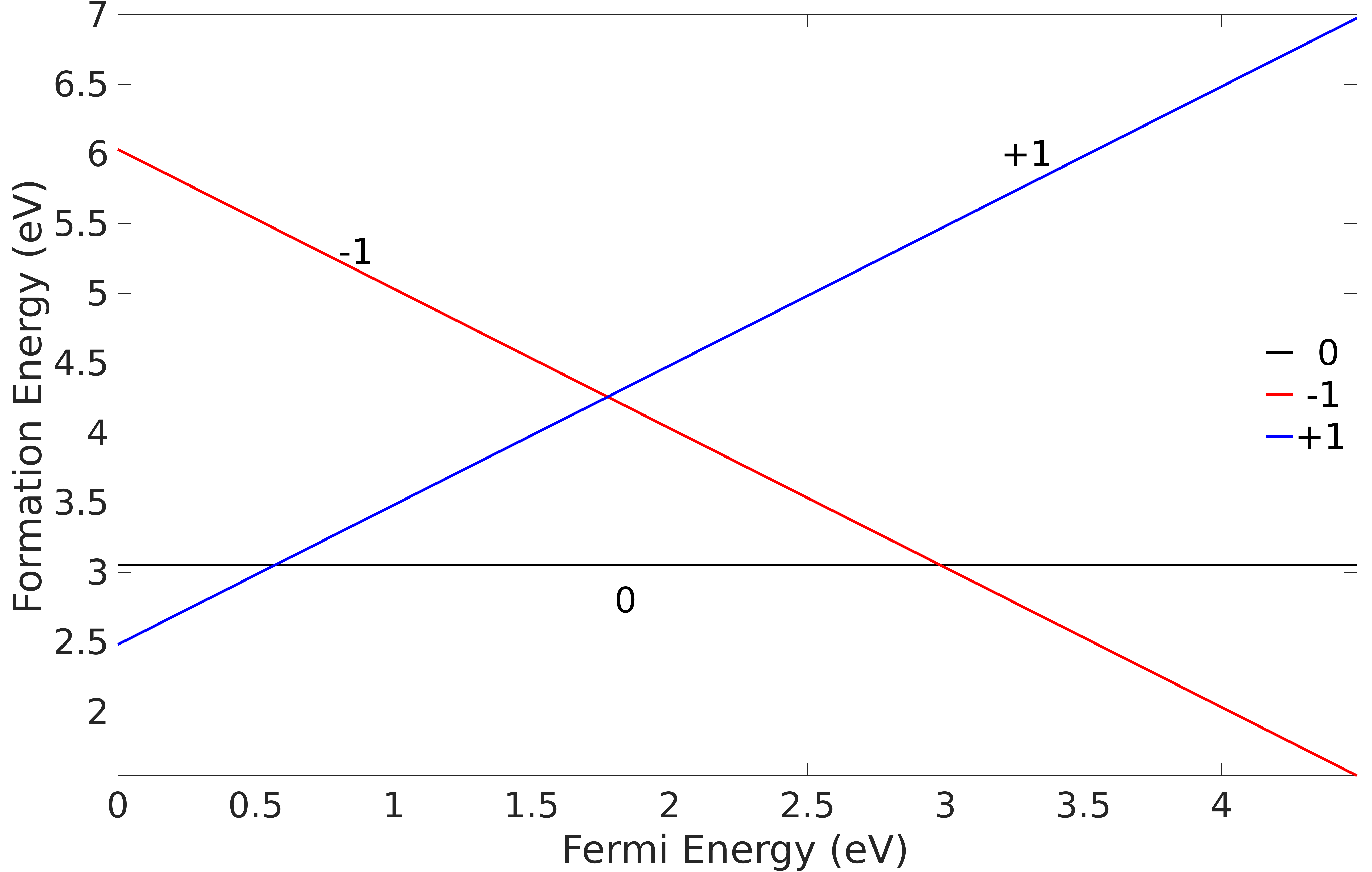}
    \caption{The formation energy of Si in boron substitution in B-rich environment.}
    \label{fig:FE-SiBB}
\end{figure}

\begin{figure}[h!]
    \centering
    \includegraphics[width=.37\textwidth]{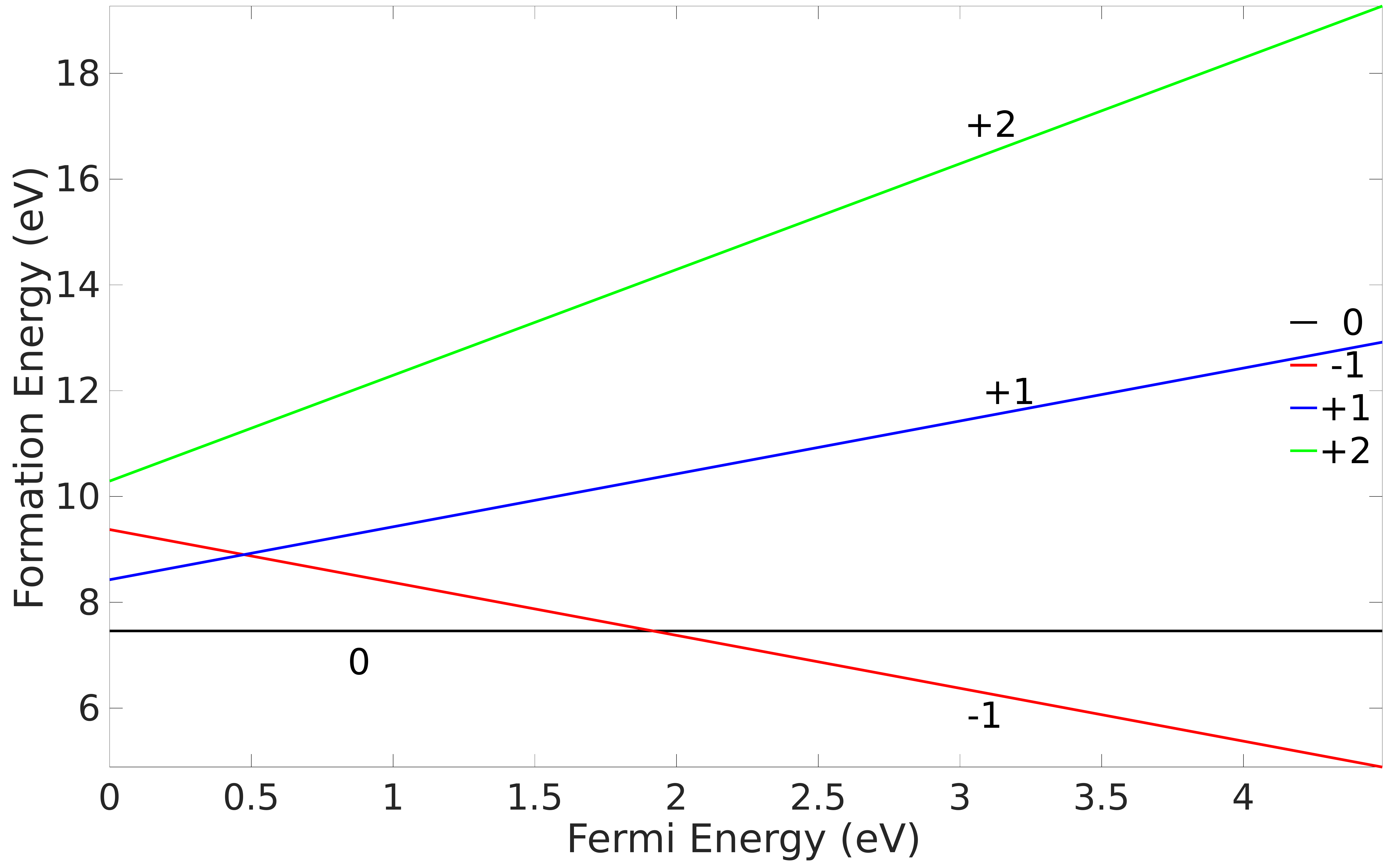}
    \caption{The formation energy of Si in nitrogen substitution in N-rich environment.}
    \label{fig:FE-SiNN}
\end{figure}

\end{document}